\documentclass[aps,superscriptaddress,pra,twocolumn,showpacs]{revtex4-1}
\usepackage{amsfonts, amsmath, amssymb, bm, bbm, graphicx, epsfig, epstopdf}
\usepackage{braket}
\usepackage{dcolumn}
\usepackage{textcomp}
\usepackage{hyperref}
\usepackage{color}
\usepackage{soul}
\usepackage[normalem]{ulem} 
\usepackage{color}

\begin{document}
	\title{Cavity-assisted controlled phase-flip gates}
	
\author{F. Kimiaee Asadi}
\email{Faezeh.kimiaeeasadi@ucalgary.ca}
\author{S. C. Wein}
\author{C. Simon}
\affiliation{Institute for Quantum Science and Technology, and Department of Physics \& Astronomy, University of Calgary, 2500 University Drive NW, Calgary, Alberta T2N 1N4, Canada}

	\begin{abstract}
	    Cavity-mediated two-qubit gates, for example between solid-state spins, are attractive for quantum network applications. We propose three schemes to implement a controlled phase-flip gate mediated by a cavity. The main advantage of all these schemes is the possibility to perform them using a cavity with high cooperativity, but not in the strong coupling regime. We calculate the fidelity of each scheme in detail, taking into account the most important realistic imperfections, and compare them to highlight the optimal conditions for each scheme. Using these results, we discuss which quantum system characteristics might favor one scheme over another.
	\end{abstract}
	\maketitle
	\section{Introduction}

 Building a global quantum network or ``quantum internet'' \cite{kimbleQI,christoph,wehner} will enable many applications such as secure communication, enhanced sensing, and distributed quantum computing. 
 Establishing a quantum network requires interfaces between stationary qubits (e.g. superconducting qubits, trapped ions, or spins in solids) and flying qubits (photons). The various quantum internet applications also require local gates between stationary qubits. For example, two-qubit gates are necessary for entanglement storage and swapping for quantum repeater protocols \cite{sangouard}, and a basic operation for generating and manipulating entangled states for quantum computation \cite{quantum-network}. So far, to perform two-qubit gates, different types of interactions including magnetic and electric dipole-dipole interactions \cite{magnetic-trap, electric, Er-Eu} and phonon-mediated couplings \cite{phonon-Lukin} have been employed.

To be efficient, interfaces between stationary and flying qubits often need cavities. It is natural to explore if cavity-assisted interactions 
can also be used to perform two-qubit gates \cite{kimble2005, lukin, feng, cavitybus, mika}. Using cavity quantum electrodynamics (QED), one can perform two-qubit phase-flip gates between qubits encoded in two modes of the electromagnetic field (photonic qubits) \cite{Zubairy}, between a quantum system and a cavity mode \cite{Biswas}, and also between two individual quantum systems (e.g., ions, atoms etc.) inside a cavity \cite{kimble2005, lukin,feng}. Of those, the latter is of great interest due to its wide-range of applications. Unlike electric and magnetic dipole-dipole interactions, cavity-mediated interactions do not require quantum systems to be very close to each other. 

Although the strong coupling regime of cavity QED has been observed for some solid-state systems such as quantum dots \cite{SC-QD1, SC-QD2, SC-QD3} and superconducting qubits \cite{SC-superconducting}, achieving a true strong coupling regime with vacuum Rabi splitting remains an outstanding challenge in other solid-state systems that are quite attractive from a quantum internet perspective. For example, rare earth ions (REIs) are attractive because of their convenient wavelengths, narrow optical transitions, and long coherence times, but have weak dipoles \cite{longdell1}. Defect centers in diamond are also attractive because of their excellent coherence (even at room temperature in the case of the NV center); however, fabricating high-quality cavities in diamond is not straightforward.

In this paper, we propose three different cavity mediated approaches to perform controlled phase-flip gates between two individual quantum systems. All of these schemes require only a high cooperativity cavity-emitter system. Therefore, even using materials or quantum emitters that are unlikely to reach the strong coupling regime, the following schemes are applicable. We calculate explicit solutions for the fidelity of these gates in detail and compare their advantages and disadvantages.

In the first scheme, we propose to perform two-qubit controlled phase-flip gates by scattering a single photon off of a cavity-qubit system. This approach has been discussed before in the context of a strong coupling regime \cite{kimble2005, rempe}, but not in the so-called `bad cavity' regime that we also consider. For the second scheme, we discuss how to use a dissipative cavity coupling to perform a controlled phase-flip gate via a virtual photon exchange. This interaction has been explored in microwave and optical systems \cite{Blais, cavitybus, lukin}, but to our knowledge, the specific details and fidelity calculations for a cavity QED phase-flip gate using this interaction are not presented in the literature. Finally, inspired by a proposed scheme in Ref. \cite{feng}, we propose a third scheme that can perform a controlled phase-flip gate between qubits with unequal optical transition frequencies using a Raman-assisted virtual photon exchange interaction. In addition, for each scheme, we provide a complete picture of the high-fidelity regime of operation that takes into account finite cavity cooperativity, and we compute each scheme's robustness to qubit decoherence and imprecise control of detunings. Moreover, we compare these three schemes using a consistent approach to highlight the advantages and disadvantage of each scheme in the context of their application to different solid-state emitters.
\vspace{-1mm}
\section{Methods}\label{sec:schemes}
\vspace{-1mm}
The starting point for each scheme is to consider a pair of individual quantum systems (A and B) placed in a cavity. For each system, we employ two of the lowest energy levels (e.g., hyperfine or Zeeman levels) to encode one qubit within states $\ket{\uparrow}$ and $\ket{\downarrow}$. Depending on the protocol we also require one or two additional excited or ground states.
In all of these schemes we require a high-cooperativity cavity $C=4g^2/\kappa \gamma \gg1$ where $\kappa$ is the cavity decay rate, $g$ is the cavity coupling rate, and $\gamma$ is decay rate of the quantum system excited state(s). 
High cooperativity is achievable in both the bad-cavity regime where $\kappa \gg g \gg \gamma$ and the strong-coupling regime where $g \gg \kappa \gg \gamma$.

\subsection{Photon scattering}  \label{ssec:cz1} 
Cavity-assisted photon scattering is one way to perform a controlled phase-flip gate between qubits in the same cavity by scattering a single photon off of the qubit-cavity system and detecting it. Although it is not necessary to detect the photon after reflection, doing so can herald the gate, which drastically improves the gate fidelity for realistic single photon sources.

Performing a phase-flip gate using this scheme has already been discussed in the strong coupling regime \cite{kimble2005,rempe,kimble2003, one-step}. Moreover, based on this scheme, a theoretical investigation of the entanglement generation has been studied \cite{heralded-gate}.
Here, for the first time we present the fidelity calculation for this gate in both bad-cavity and strong-coupling regimes. We also analyze infidelity due to possible spectral wandering of the incident single photon and imperfect quantum systems resonance conditions.

\begin{figure}
\centering
	\includegraphics[width=7cm]{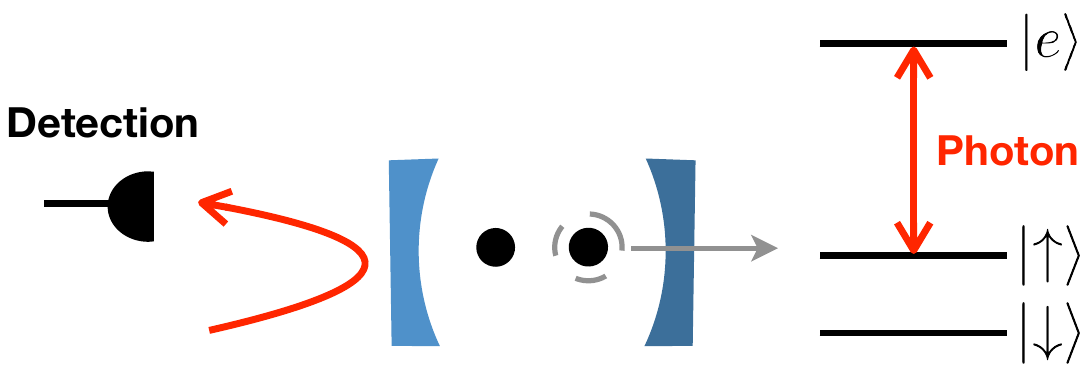}
	\caption{Scheme illustration and energy level diagram of both quantum systems in the photon scattering scheme. The $\ket{\uparrow}$ and $\ket{\downarrow}$ ground states represent the qubit states and $\ket{e}$ is the excited state of the system. The $\ket{\uparrow}$--$\ket{e}$ transition of both systems is resonant with the cavity whereas the $\ket{\downarrow}$--$\ket{e}$ transition is either far-detuned or uncoupled from the cavity.}\label{fig:cz1}
\end{figure}

In this scheme, we use a single sided cavity and two 3-state quantum systems with a $\Lambda$ system (i.e., two ground states $\ket{\uparrow}$ and $\ket{\downarrow}$ and an excited state $\ket{e}$). For both quantum systems, the $\ket{\uparrow}$--$\ket{e}$ transition is resonant with the cavity and the $\ket{\downarrow}$--$\ket{e}$ transition does not interact with the cavity, as shown in Fig.~\ref{fig:cz1}.
In systems where both qubit states can interact with the excited state for the same polarization (e.g., rare earth ions), $\ket{\downarrow}$--$\ket{e}$ should be far-detuned from the cavity frequency \cite{Obrien}.

We denote the state of the photon by $\ket{p}$. If both qubits are in the state $\ket{\downarrow}$, the photon enters and then exits the cavity unhindered. This reflection of the photon from inside the cavity causes the joint state of the qubit-photon system to acquire a $\pi$-phase shift. 
On the other hand, if either or both of the qubits are in the state $\ket{\uparrow}$, the cavity mode becomes modified and the photon will not be impedance matched. In this case, the cavity acts as a mirror and the photon does not enter the cavity but reflects from the out-coupling mirror of the cavity. Under the correct cavity and photon conditions, the phase of the qubit-photon system remains unchanged for these three cases.

This phase-flip gate can also be described by a unitary operator
$U=e^{i\pi\ket{\downarrow\downarrow}\bra{\downarrow\downarrow}\otimes \ket{p}\bra{p}}$, meaning that there would be a phase-flip in the system only if both ions are in the state $\ket{\downarrow}$.
As a result the states $\ket{\uparrow\uparrow}\ket{p}$, $\ket{\uparrow\downarrow}\ket{p}$ and $\ket{\downarrow\uparrow}\ket{p}$ remain unchanged but $\ket{\downarrow\downarrow}\ket{p}$ changes to $-\ket{\downarrow\downarrow}\ket{p}$. At the end, we can detect the reflected photon to herald the gate.

In the strong coupling regime, the impedance mismatch can be described simply by a frequency shift (vacuum Rabi splitting). However, in the bad cavity regime, the resonant systems cause a phase shift that destroys the constructive interference of the photon inside the cavity within a narrow frequency window; therefore, the photon will not enter the cavity (see ref. \cite{Obrien}).

In the regime where $C\gg 1$ we find that the total gate fidelity of this scheme is well-approximated by
\begin{equation}
\label{eq:kimble}
\begin{aligned}
F_\text{gate}= 1&-\frac{5}{4C}-\frac{\delta_p^2+\sigma_p^2}{8 \gamma^2C^2}\!\left[11-20\left(\frac{2g}{\kappa}\right)^2\!+12\left(\frac{2g}{\kappa}\right)^4\right]\\
&-\frac{\left(\delta_{\epsilon_A}\!-\delta_{\epsilon_B}\right)^2}{4\gamma^2C}-\Gamma T,
\end{aligned}
\end{equation}
 where $\sigma_p$ is the spectral standard deviation of an incident photon with a Gaussian intensity profile, $\delta_p$ is the mean cavity-photon detuning, and $\delta_{{\epsilon}_k}$ for \mbox{$k\in\{A,B\}$} is the detuning of the $k^\text{th}$ system's optical transition from the cavity resonance. We also introduce $\Gamma$ as the effective qubit decoherence rate that is a weighted average of decoherence rates from system-specific processes that are small compared to cavity dissipation and spontaneous emission. Here $T=8\pi\sqrt{2\ln{2}}/\sigma_p$ is the gate time, which we define to be twice the FWHM duration of the photon for this scheme. This effective rate is at least given by the qubit decoherence time: $\Gamma \geq 1/2T_2$. Equation (\ref{eq:kimble}) is valid to first order in $C^{-1}$ and $\Gamma T$; and to second order in $\delta_{\epsilon_k}/\gamma$, $\delta_p/\gamma C$, and $\sigma_p/\gamma C$. See Appendices \ref{Approach} and \ref{AppendixB} for detailed calculations.

\begin{figure}[t]
\centering
     \begin{flushleft}\hspace{0mm}{(a)}\end{flushleft}\vspace{-5mm}
 	\mbox{\hspace{-5mm}\includegraphics[width=7.8cm]{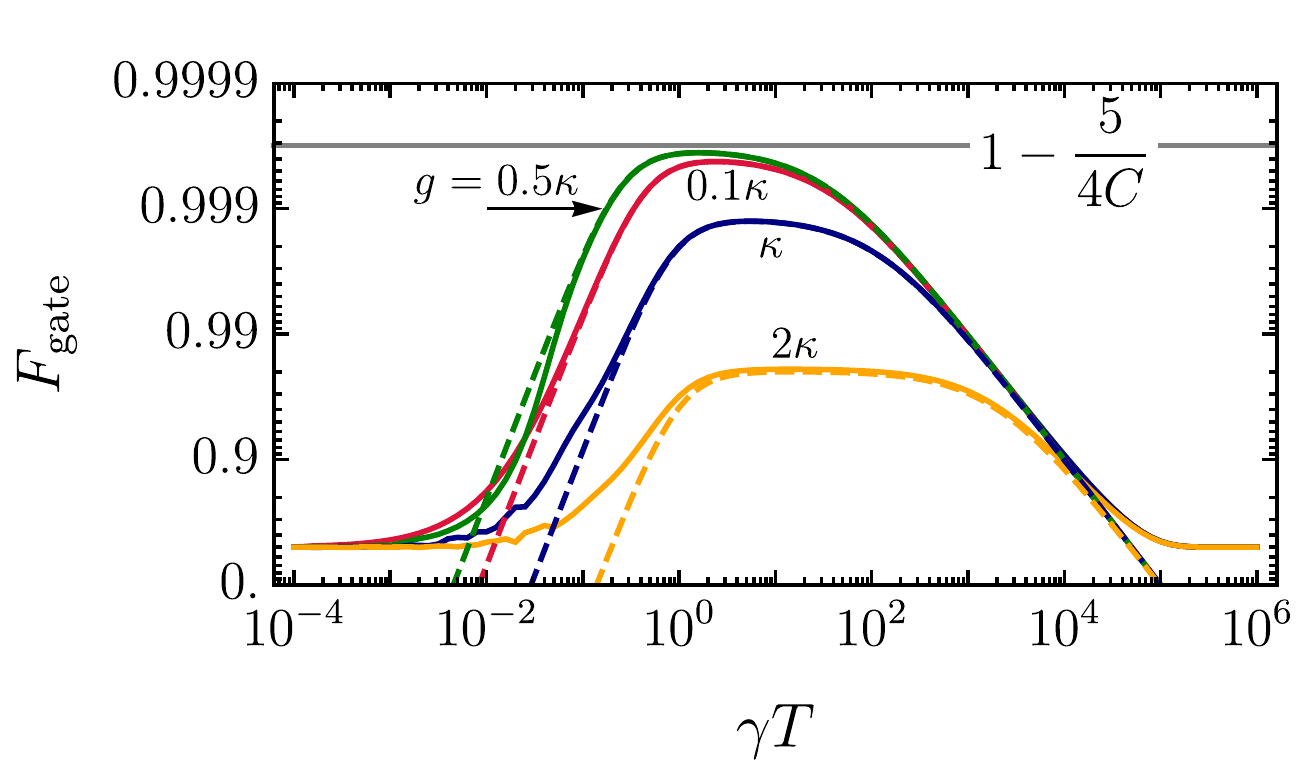}}\vspace{-8mm}
     \begin{flushleft}\hspace{0mm}{(b)}\end{flushleft}\vspace{-5mm}
 	\mbox{\hspace{-5mm}\includegraphics[width=7.8cm]{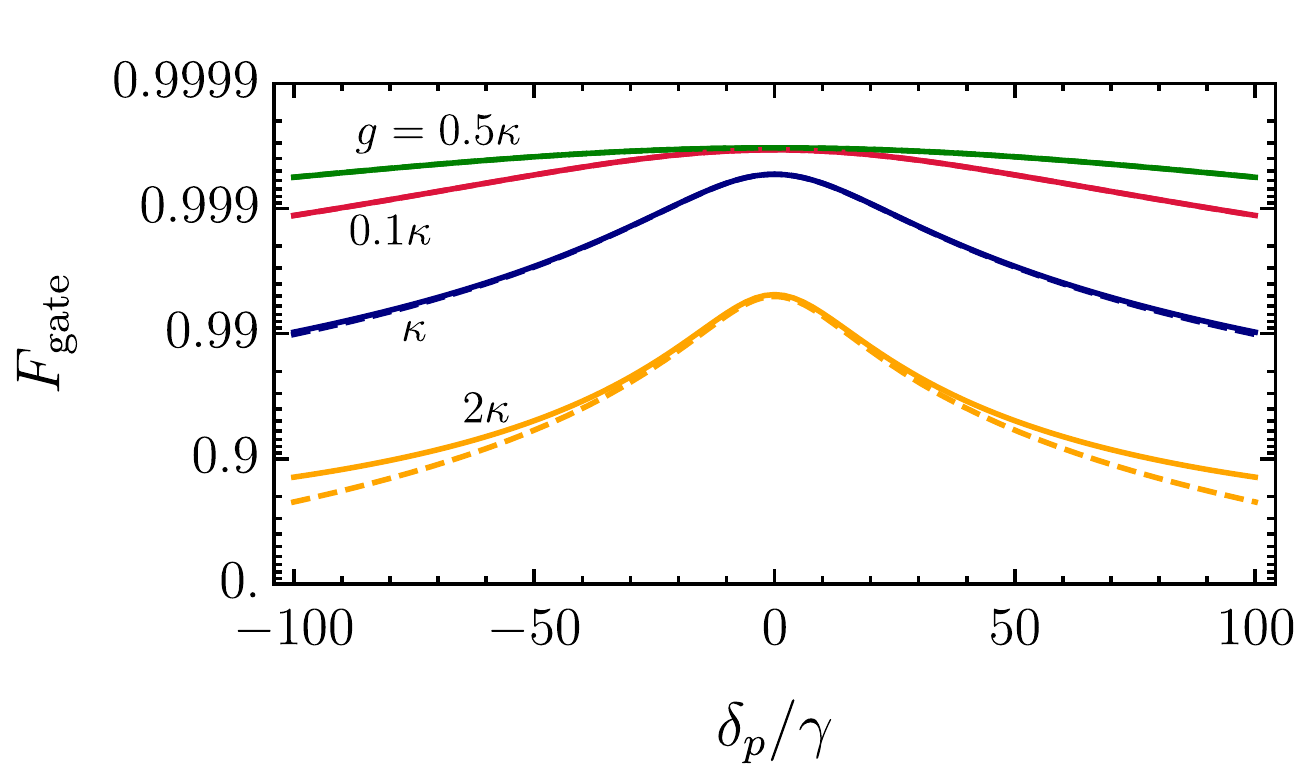}}\vspace{-8mm}
 	\begin{flushleft}\hspace{0mm}{(c)}\end{flushleft}\vspace{-5mm}
 	\mbox{\hspace{-5mm}\includegraphics[width=7.8cm]{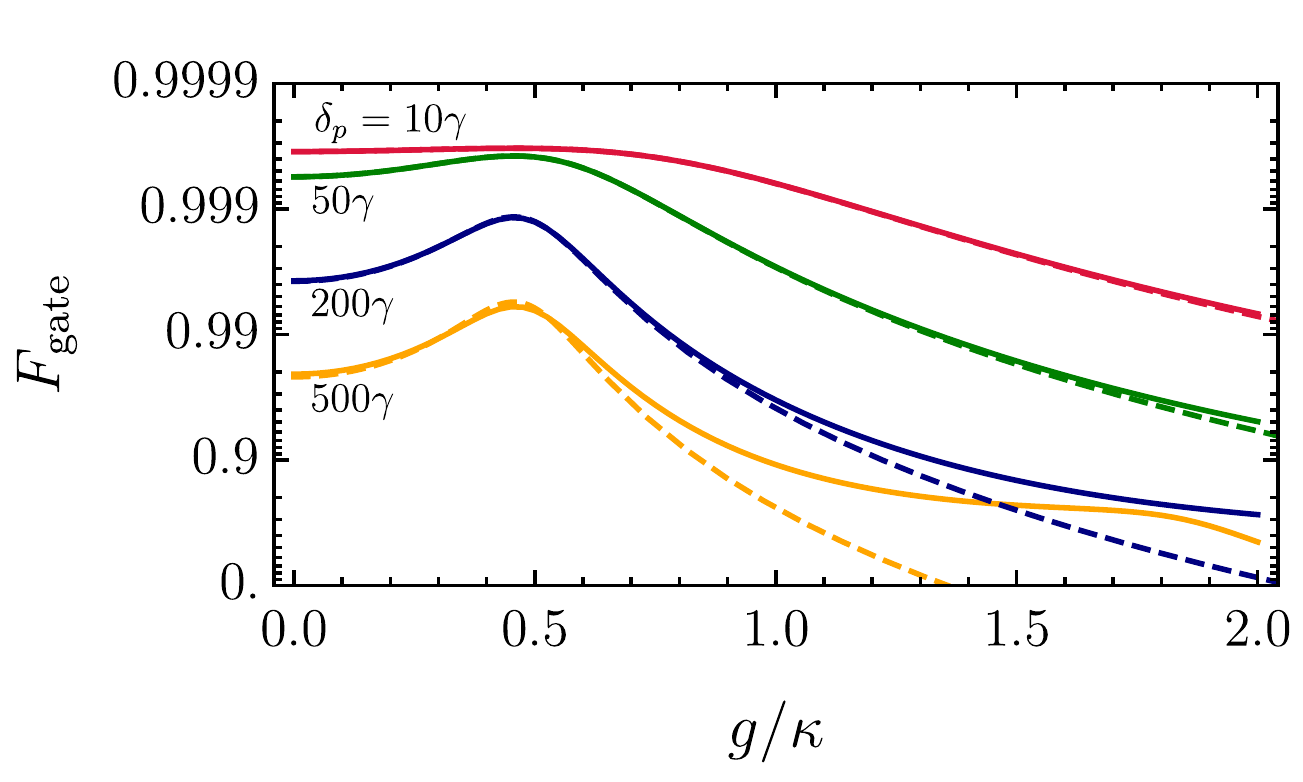}}
   
	\caption{Fidelity of the photon scattering phase gate as a function of (a) the gate time $T$, (b) the photon-cavity detuning $\delta_p$ for a given cavity regime, and (c) the cavity regime for a given photon-cavity detuning. Here we set the cavity cooperativity to $C=4000$, quantum system detunings to $\delta_{\epsilon_A}=\delta_{\epsilon_B}=0$, the effective dephasing to $\Gamma/\gamma=10^{-5}$, and for (a): $\delta_p=30\gamma$, and (b, c): $T=2/\gamma$. The ratio of cavity coupling rate $g$ to cavity dissipation rate $\kappa$ gives the regime of operation with $g/\kappa \ll 1$ and $g/\kappa\gg 1$ indicating the bad-cavity and strong-coupling regimes, respectively. The solid lines show the numerical solution without expanding around small $\delta_p/\gamma C$, $\sigma_p/\gamma C$, and large $C$. The dashed lines correspond to the analytic approximation given by equation (\ref{eq:kimble}). For the numerical simulation, we assume that the effective decoherence rate $\Gamma$ exponentially degrades the coherence of the reduced spin density matrix as opposed to using the linear correction as in the analytic approximation.}
	\label{fig:F-scattering}
\end{figure}

The second term in equation (\ref{eq:kimble}) shows the infidelity due solely to a finite cavity cooperativity,  the third term shows the spectral mode matching sensitivity of the incident single photon, the fourth term captures the degradation when either of the quantum systems $\ket{\uparrow}$ transitions are not exactly resonant with the cavity, and the last term captures a linear scaling due to a small effective qubit decoherence rate.

For shorter photons, the larger bandwidth can exceed the narrow spectral range over which the destructive interference occurs within the cavity. This degrades the fidelity of the phase-flip gate. On the other hand, for very long photons, the gate becomes so slow that qubit decoherence can dominate and degrade the gate fidelity. These two competing processes limit high-fidelity operation to a range of gate times (see Fig.~\ref{fig:F-scattering}.a) with an optimal gate time given by $T_o^3 = (352\pi^2\ln{2})/(\gamma^2C^2\Gamma)$ in the bad-cavity limit.

We find that there is a non-trivial relationship between the regime of operation and robustness against photon detuning (see Fig.~\ref{fig:F-scattering}.b and \ref{fig:F-scattering}.c). In addition, since the photon detuning infidelity scales as $\delta_p^2$, averaging the fidelity over a Gaussian spectral wandering profile with standard deviation $\sigma^\star$ simply results in $\sigma^\star$ replacing $\delta_p$ in equation (\ref{eq:kimble}). Hence, the effect of photon detuning captured in equation (\ref{eq:kimble}) and shown in Fig.~\ref{fig:F-scattering} also demonstrates the effect of random spectral wandering of the incident photon.

Far in the bad cavity regime where $g/\kappa\ll 1$, the system is very resilient to photon detuning and finite bandwidth effects. In this regime, the spectral window where a phase flip can occur is small, scaling by $g^2/\kappa$ \cite{Obrien}; however, mode matching so that the photon enters the cavity when the quantum systems are off resonant is relatively simple to accomplish due to the large cavity bandwidth. In contrast, in the strong-coupling regime where $g/\kappa\gg 1$, the spectral window for the phase flip is larger, but mode matching becomes much more difficult to achieve and so the fidelity is more sensitive to the photon spectral properties. Although the bad-cavity regime is surprisingly robust against spectral wandering, photon detuning, and finite bandwidth effects, we find that the most robust regime is the so-called `critical regime' where $2g\simeq \kappa$ \cite{criticalregime} (see Fig.~\ref{fig:F-scattering}.b).

\subsection{Simple virtual photon exchange} \label{ssec:cz2} 
Another type of a cavity-assisted interaction between qubits can be achieved by the exchange of virtual cavity photons when the quantum systems' optical transitions are resonant but dispersively coupled to a cavity mode \cite{Blais, cavitybus, lukin}.
Using this interaction, we provide a description of how to perform a phase-flip gate as well as detailed calculations on the fidelity of the gate. 

\begin{figure}[t]
\centering
	\includegraphics[width=8.5cm]{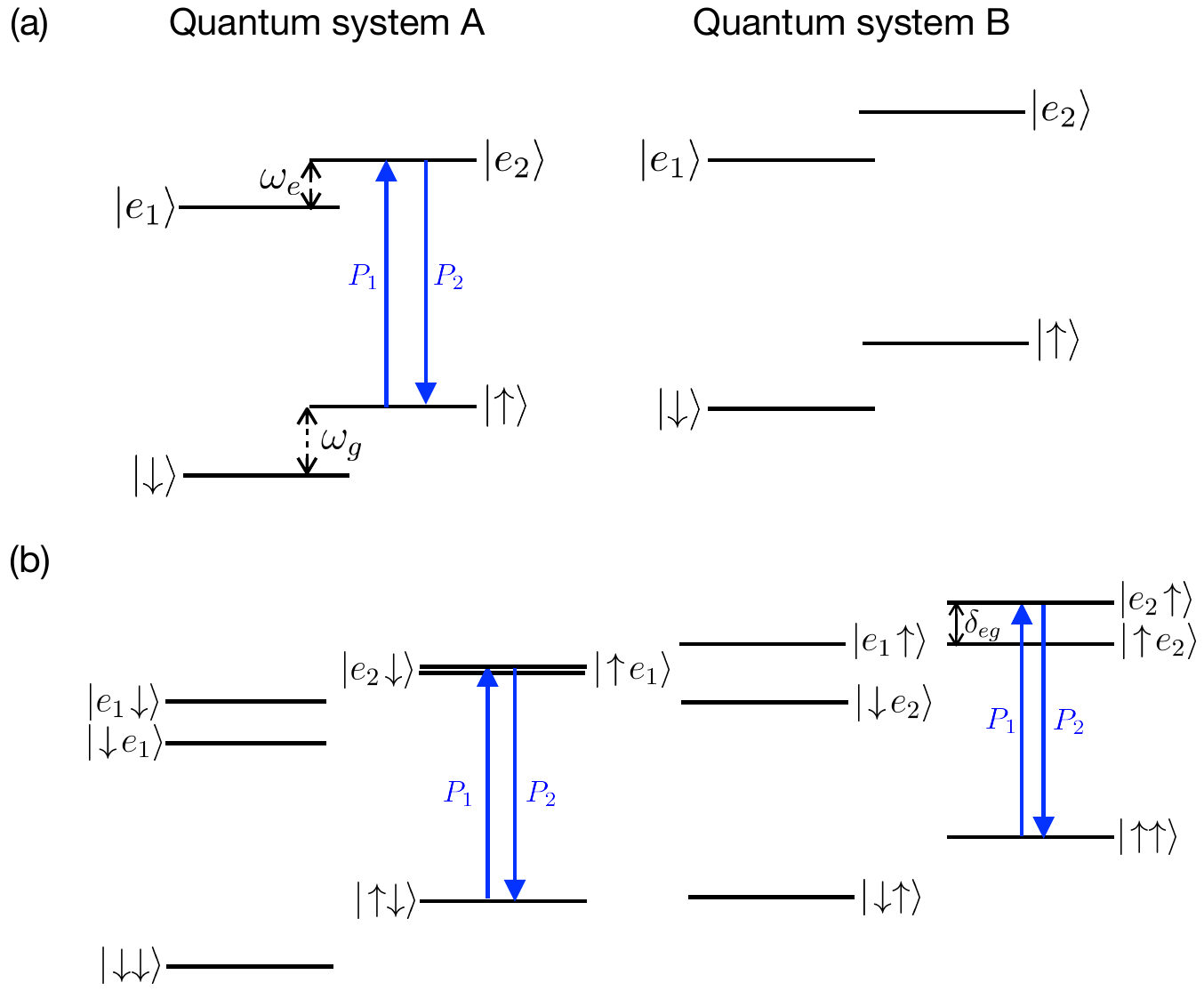}
	\caption{(a) Energy level diagram of the simple virtual photon exchange phase gate.  The $\ket{\uparrow}$--$\ket{e_2}$ transition of system A is brought into resonance with the $\ket{\downarrow}$--$\ket{e_1}$ transition of system B. To perform a phase-flip gate, we apply a pair of optical $\pi$ pulses with a time delay on system A. b) Level diagram in the product space. The splitting between $\ket{e_2\uparrow}$ and $\ket{\uparrow e_2}$ states is equal to the difference between the ground and excited states splittings $\delta_{eg}=|\omega_e-\omega_g|$. The interaction between the states $\ket{e_2\downarrow}$ and $\ket{\uparrow e_1}$ performs a phase flip gate in the system. However, high fidelity can only be achieved when $\delta_{eg}$ is large enough so that the $\ket{e_2\uparrow}$ and $\ket{\uparrow e_2}$ states are not interacting.} \label{fig:cz2}
\end{figure}

For our analysis, we consider two 4-state quantum systems; each system has two ground states $\ket{\downarrow}$ and $\ket{\uparrow}$ and two excited states $\ket{e_1}$ and $\ket{e_2}$. Since the systems' optical transitions are dispersively coupled to a symmetric cavity, there is no energy exchange with the cavity. To have a phase-flip gate between qubits, first we bring the $\ket{\uparrow}$--$\ket{e_2}$ transition of the first system into resonance with the $\ket{\downarrow}$--$\ket{e_1}$ transition of the second system using a magnetic flux or an AC Stark pulse \cite{cavitybus, stark}. Next, a $\pi$-pulse ($P_1$) is applied to one of the systems to bring it to the excited state, as shown in Fig.~\ref{fig:cz2}.a. After a time delay, another optical $\pi$-pulse ($P_2$) is applied to bring the excited quantum system back to its initial state. 

To understand how this process performs a phase flip gate between qubits, we have shown the level diagram of the two-qubit system in the product space in Fig.~\ref{fig:cz2}.b. If the two-qubit system is in the state $\ket{\downarrow\downarrow}$ or $\ket{\downarrow\uparrow}$, pulses $P_1$ and $P_2$ are ineffective and hence the qubits will be unaffected. If the state is $\ket{\uparrow\uparrow}$, then $P_1$ excites system A to the excited state $\ket{e_2}$. So far as the splitting between the states $\ket{e_2\uparrow}$ and $\ket{\uparrow e_2}$ is large enough, there will be no interaction between them. Applying another $\pi$-pulse $P_2$ to system A, will then return it to the ground state and leave $\ket{\uparrow\uparrow}$ unaffected. However, if the state is $\ket{\uparrow\downarrow}$, after exciting system A to the excited state, the degenerate states $\ket{e_2\downarrow}$ and $\ket{\uparrow e_1}$ interact via the virtual exchange of a cavity photon which adiabatically performs a $\pi$ phase flip on the state. At the end, the optical pulse $P_2$ brings system A back to its initial state but with a relative phase $-\ket{\uparrow\downarrow}$.

The virtual interaction can be controlled by detuning the quantum system optical frequencies away from each other. That is, when the detuning between the optical transitions is much more than the cavity coupling strength, the qubit-qubit interaction can be made negligible and the gate will not be successful.

\begin{figure}
\centering
	\mbox{\hspace{-9mm}\includegraphics[width=7.5cm]{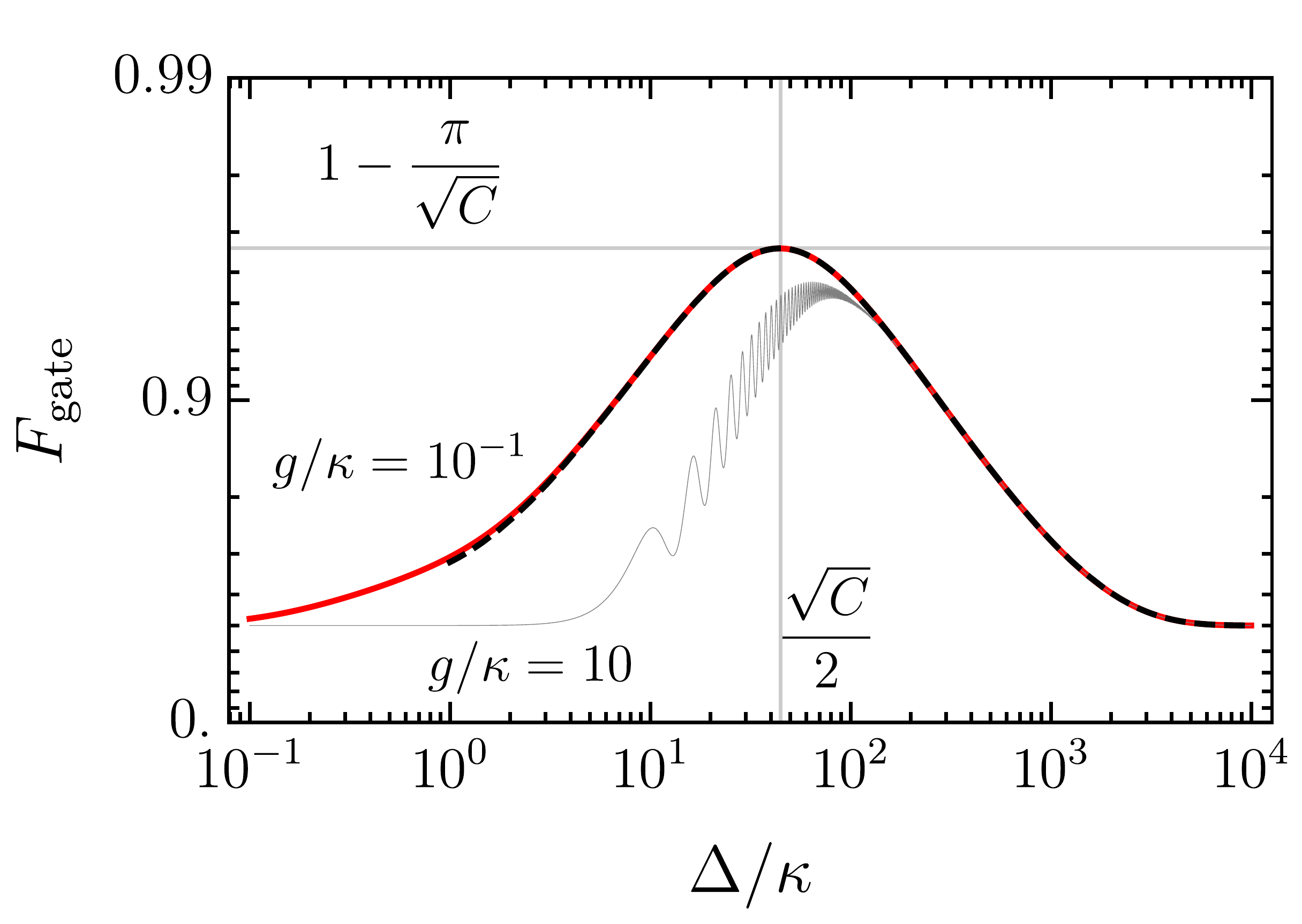}}
	\caption{Fidelity of the simple virtual photon exchange phase gate $F_\text{gate}$ as a function of the cavity detuning $\Delta/\kappa$. The analytic solution in the adiabatic regime given in Appendix B (black dashed curve) matches closely with the numerically exact solution in the weak coupling regime (solid red curve) with $g/\kappa=10^{-1}$ and accurately predicts the maximum fidelity $F_\text{max}$ at $2\Delta=\kappa\sqrt{C}$. High fidelity can also be achieved if the system is not too far into the strong-coupling regime (gray solid curve) with $g/\kappa=10$, but it is less optimal. Here $\Delta_\epsilon=0$, $\delta_{eg}\rightarrow\infty$, and $C=8000$.}\label{fig:virtualphoton}
\end{figure}

We define $\Delta_k = \omega_C - \omega_k$ as the detuning between the cavity and the $\ket{\uparrow}\rightarrow\ket{e_2}$ transition of the $k^\text{th}$ quantum system.
In the high cooperativity regime, the cavity detuning \mbox{$\Delta=\Delta_A\simeq\Delta_B+\delta_{eg}$} dictates the gate fidelity $F_\text{gate}$ and gate time $T=\pi\Delta/g^2$, where $\delta_{eg}=|\omega_\text{e}-\omega_\text{g}|$ is the difference between ground-state and excited state splittings. If the detuning is too small, the excited system can emit a photon into the cavity mode, which can subsequently decay. Hence the fidelity becomes limited by the cavity dissipation rate $\kappa$. However, if the detuning is too large, $T$ is also large, which causes the system to relax before the gate is complete. Hence the fidelity becomes limited by $\gamma$. The maximum gate fidelity of the simple virtual photon exchange is achieved between these extremes at a detuning of $2\Delta=\kappa\sqrt{C}$ (see Fig.~\ref{fig:virtualphoton}) and is well-approximated by
\begin{equation}
\label{fid2}
    F_\text{max} = 1 - \frac{\pi}{\sqrt{C}}-\frac{3\pi^2}{32}\left[\left(\!\frac{T_o\Delta_\epsilon}{2\pi}\!\right)^2\!\!+\left(\!\frac{2\pi}{T_o\delta_{eg}}\!\right)^2\!\!-\frac{12}{C}\right]-\Gamma T_o,
\end{equation}
where $T_o=2\pi/\gamma\sqrt{C}$ is the optimal gate time, $\Gamma$ is the effective decoherence rate, and $\Delta_{\epsilon}=\omega_B-\omega_A-\delta_{eg}$ is a small detuning between the systems' optical transitions. Equation (\ref{fid2}) is valid to first order in $C^{-1}$ and $\Gamma T_o$, and also to second order in $T_o\Delta_\epsilon$ and $(T_o\delta_{eg})^{-1}$. See Appendices \ref{Approach} and \ref{AppendixB} for detailed calculations and the expression for $F_\text{gate}$ that includes dependence on $\Delta$ and $\kappa$.

From the above solution we can note that the maximum fidelity is ultimately limited by the cavity cooperativity. However, we can also see that this maximum can only be reached if the optical transition of the systems are resonant to within a precision dictated by the inverse gate time: $\Delta_\epsilon \ll 2\pi T_o^{-1}$. In addition, there should not be any other optical transitions coupled to the cavity within $\delta_{eg}\gg 2\pi T_o^{-1}$. If either or both of these conditions are violated, it may be beneficial to choose a detuning that better optimizes the gate fidelity.

This simple virtual photon exchange scheme operates most optimally in the bad-cavity regime. In the strong-coupling regime, Rabi oscillations begin to occur when the cavity detuning is not large enough. This effect pushes the optimal detuning further away and forces the fidelity to be more limited by decay from the excited state (see Fig.~\ref{fig:virtualphoton}).

\subsection{Raman virtual photon exchange}  \label{ssec:cz3}
A controlled phase-flip gate can also be performed between distant qubits by virtual excitation of the cavity mode via a Raman coupling. Performing two-qubit gates using the Raman coupling has been discussed in Ref.~\cite{imamoglu} for quantum dots. Later, Ref.~\cite{feng} proposed an improved version of the latter scheme for trapped ions, which is more efficient in terms of the number of operations. However, there is a challenge related to shelving the qubit state in Ref.~\cite{feng} (see below for more information). Here, we discuss and fully analyze our modified scheme that overcomes that challenge without limiting our analysis to a specific system. Using our proposed scheme, one may perform a controlled phase-flip gate between qubits in quantum systems with unequal optical transitions using a two-photon resonance between a driving laser and a vacuum cavity field.

\begin{figure}
\centering
	\includegraphics[width=8.5cm]{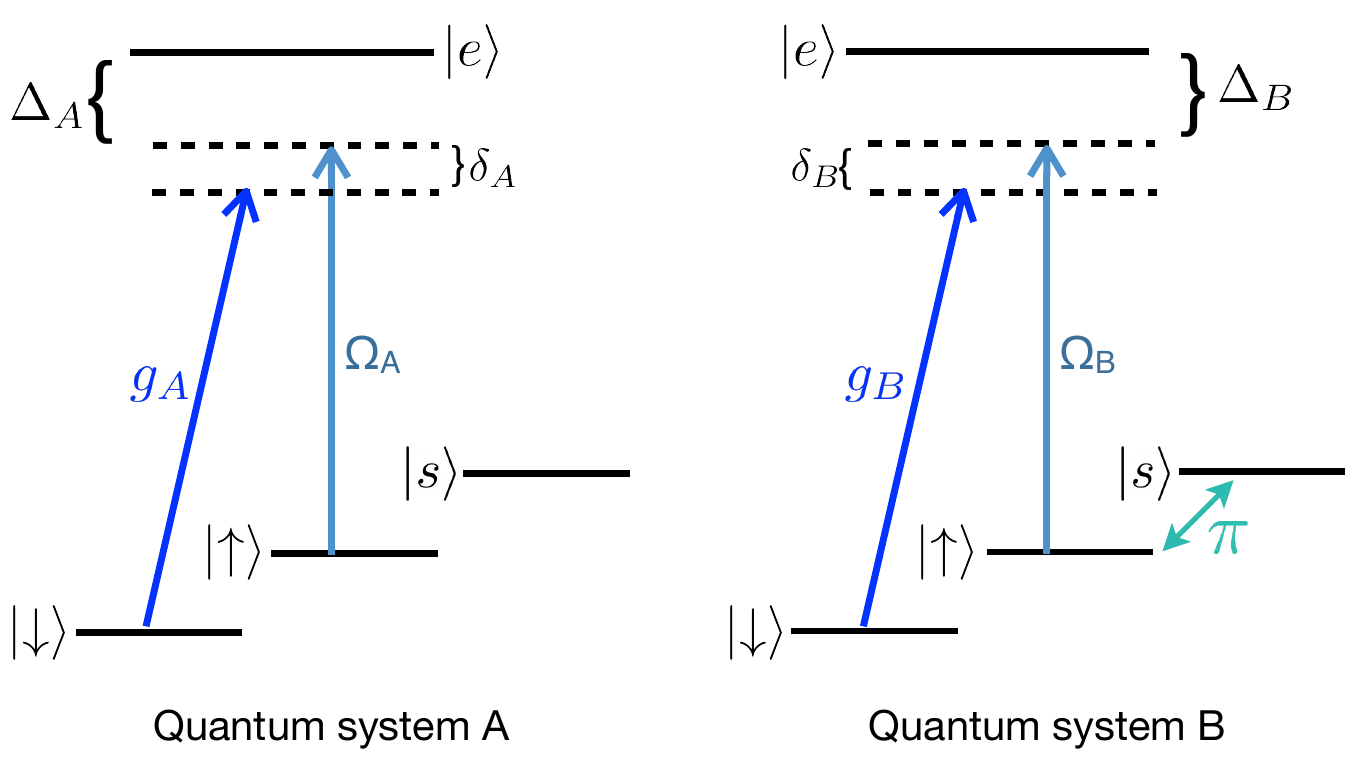}
	\caption{Energy-level diagrams and pulse sequence for the Raman virtual photon exchange phase gate. For each system, we establish the near two-photon resonance process using a classical laser field with Rabi frequency $\Omega_k$ and the cavity-mode with cavity coupling rate $g_k$, for $k\in\{A,B\}$.  A MW $\pi$ pulse shelves qubit B into the state $\ket{s}$. Then, laser A (B) applies a $2\pi$ pulse on qubit A (B). Another MW $\pi$ pulse then brings qubit B back to its original state.}\label{fig:cz3}
\end{figure}

For our analysis, we consider two 4-level quantum systems each containing three ground states that includes two qubit states $\ket{\uparrow}$ and $\ket{\downarrow}$ and a shelving state $\ket{s}$ in addition to an excited state $\ket{e}$. The systems are dispersively coupled to a far-detuned cavity with a high cooperativity. For each system, we drive the Raman transition between qubit ground states via the vacuum cavity field and a driving field with Rabi frequencies $g_k$ and $\Omega_k$, respectively, for system $k\in\{A,B\}$ as shown in Fig.~\ref{fig:cz3}. The detuning $\Delta_k$ is assumed to be large compared to the Rabi frequencies $\Omega_k$ so that the excited state will not be populated by the driving fields. For a fixed cavity frequency $\omega_c$, if the driving fields are tuned to satisfy the resonance condition $\delta_A=\delta_B=\delta$, then an effective coupling between $\ket{\uparrow\downarrow}$ and $\ket{\downarrow\uparrow}$ is induced.

To perform the gate, first a microwave (MW) $\pi$ pulse is applied to shelve qubit B (i.e., target qubit) to the shelving state $\ket{s}$. 
Next, the driving fields A and B are turned on to induce the Raman coupling. During the adiabatic Raman process, the qubits interact through the cavity mode via a virtual photon interaction. Once the Raman process is complete, another MW $\pi$ pulse brings qubit $B$ back to the state $\ket{\uparrow}$.
The result of this gate transforms $\ket{\uparrow\downarrow}$ into $-\ket{\uparrow\downarrow}$ without affecting the relative phases of the remaining qubit product states.

In the following, we assume that $g_A=g_B=g$; however, we discuss how to compensate for unequal cavity couplings in Appendix B. For high fidelity operation, it is necessary to satisfy four main conditions: (1) the two-photon resonance detuning $\delta$ must be larger than the cavity linewidth $\kappa$, (2) the gate time given by $T=\pi\delta\Delta_A\Delta_B/g^2\Omega_A\Omega_B$ must not exceed the lifetime of the shelved state, (3) the driving field intensities should not exceed any detunings $\Omega_k\ll\Delta_k,\delta$, and (4) the system should be in the bad-cavity regime $g<\kappa$.

The maximum gate fidelity of the Raman virtual photon exchange in the high cooperativity regime is achieved under the condition that $2\delta=\kappa\sqrt{C}$, and is well approximated by
\begin{equation}
\label{fidelity-feng}
    \!F_\text{max}\!= 1 - \frac{\pi}{\sqrt{C}}-\frac{\pi^2}{16}\left[\left(\!\frac{T_o\delta_\epsilon}{2\pi}\!\right)^2\!+\frac{\Delta_\epsilon^2}{\Delta^2}-\frac{18}{C}\right]-\Gamma T_o,
\end{equation}
where $T_o = (\Delta/\Omega)^2(2\pi/\gamma\sqrt{C})$ is the optimal gate time, $\Omega=\Omega_A=\Omega_B\sqrt{\Delta_B/\Delta_A}\simeq\Omega_B$ is the optimal Rabi frequency condition, $\Delta_k$ is the detuning between the $k^\text{th}$ quantum system's optical transition and the driving field, $\Gamma$ is the effective decoherence rate that includes decoherence caused by the shelving state decay rate $\gamma_s\ll \gamma$, and $\delta_{\epsilon}=|\delta_A-\delta_B|\ll \delta$ is a small two-photon resonance error. This expression is valid to first order in $C^{-1}$ and $\Gamma T_o$, and also to second order in $T_o\delta_\epsilon$ and $\Delta_\epsilon/\Delta=|\Delta_A-\Delta_B|/\Delta$, where $\Delta=(\Delta_A+\Delta_B)/2\simeq\Delta_A\simeq\Delta_B$. See Appendices \ref{Approach} and \ref{AppendixB} for detailed calculations and the full expression for $F_\text{gate}$ that includes the dependence on $\delta$ and $\kappa$.

The Raman scheme exchange is slower than the simple exchange by a factor of $(\Omega/\Delta)^{-2}\gg 1$, which must be large so that the excited states $\ket{e}_k$ are only virtually populated. Although the gate is slower, it gains the advantage of being much less sensitive to $\gamma$. In fact, these two factors seem to exactly cancel to give the same cooperativity scaling. However, as a consequence of the Raman scheme being slower, it can also suffer a lot from the decay of $\ket{s}$. Therefore, for high fidelity operation, it is necessary to shelve $\ket{\downarrow}_B$ into a metastable state where $\gamma_s \ll \gamma$ so that $\gamma_s/\gamma \ll (\Omega/\Delta)^2\ll 1$. In this regime, the first-order correction due to the shelving decoherence is $-\gamma_s T/8$. Thus the effective decoherence rate $\Gamma$ must include at least a contribution of $\gamma_s/8$.

\begin{figure}[t]
\centering
    \mbox{\hspace{-5mm}\includegraphics[width=7.5cm]{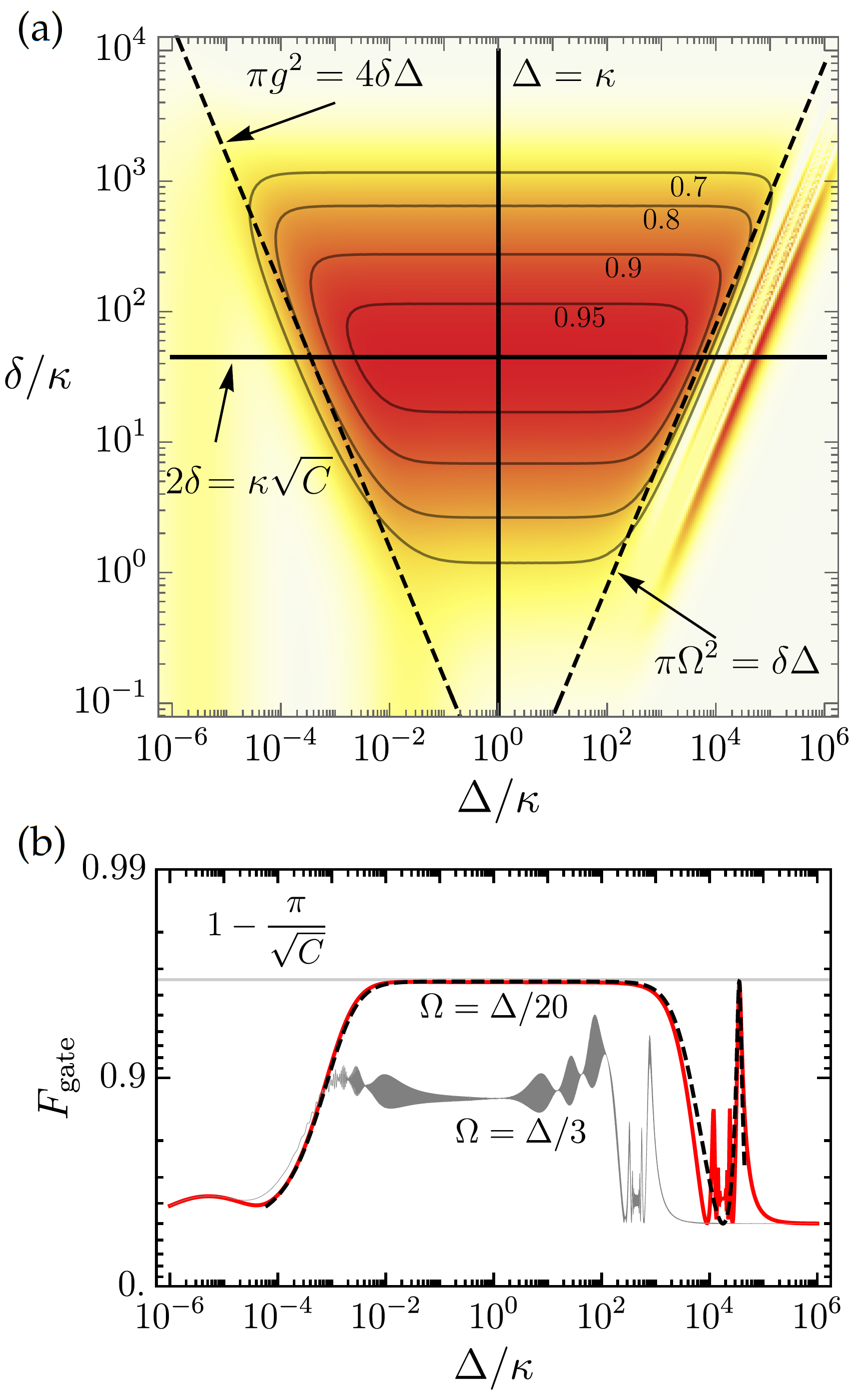}}
	\caption{(a) Numerically-simulated fidelity of the Raman-induced virtual photon exchange phase gate $F_\text{gate}$ as a function of detunings $\delta$ and $\Delta$ for a fixed $\Omega = \Delta/20$ and $g/\kappa = 10^{-1}$. The analytic bounds on the regime of high fidelity are marked with dashed black lines. The cross-section along $\Delta=\kappa$ is identical to the fidelity curve plotted in figure \ref{fig:virtualphoton}; but in the Raman scheme, $\delta$ has the same function as the cavity detuning in the simple exchange scheme. The maximum fidelity occurs along the ridge $2\delta=\kappa\sqrt{C}$ (solid black horizontal line). (b) Cross-section along the $2\delta=\kappa\sqrt{C}$ line. The fidelity oscillates rapidly for larger values of $\Omega/\Delta$ (gray solid curve) as the driving laser begins to induce coherence between the ground and excited states and the adiabatic evolution breaks down. The red and gray lines are numerically computed while the black dashed line shows the analytic solution detailed in the appendix. Other parameters are $\delta_\epsilon=0$, $\Delta_A=\Delta_B$, $\Gamma=0$, and $C=8000$.}\label{fig:ramanscheme}
\end{figure}
Similar to the simple virtual photon scheme, the Raman scheme has a fairly strict resonance condition scaling by $T_o^{-1}$. That is, for high fidelity operation, it is necessary that the two-photon resonance be satisfied more precisely than the inverse gate time: $\delta_\epsilon\ll 2\pi T_o^{-1}$.

The main advantage of the Raman scheme is that the gate fidelity is robust against unequal optical transitions. Since the maximum fidelity depends only on the relative detuning difference $\Delta_\epsilon=|\Delta_A\!-\!\Delta_B|$ compared to the magnitude $|\Delta|$, there is an inherent trade off between gate time and system spectral separation. The larger the difference between the system transitions, the larger both cavity detunings must be to maintain the same fidelity. This, in turn, increases the overall optimal gate time. However, $\Delta$ cannot be increased indefinitely because the fidelity will eventually become limited by decoherence. By considering the bounds on the regime of high-fidelity (see figure \ref{fig:ramanscheme}), we find that the spectral separation that will give a maximum fidelity no less than $1-2\pi/\sqrt{C}$ is $\Delta_\epsilon=\kappa\gamma/(\pi\Gamma\sqrt{8})$, which corresponds to $\Omega=2\Delta\sqrt{\Gamma/\gamma}$ and $2\Delta=\Delta_\epsilon \sqrt{\pi\sqrt{C}}$ (see Appendix \ref{AppendixB}). This limit on $\Delta_\epsilon$ implies that, for $\Gamma/\gamma\ll 1$, the spectral separation of the systems can be many times larger than the cavity linewidth without significantly degrading the fidelity.

Let us note that in Ref. \cite{feng} the state $\ket{\downarrow}$ of one qubit is shelved in the excited state $\ket{e}$. Doing so causes an additional unwanted phase evolution on the shelved state due cavity Lamb and AC Stark shifts that cannot be reduced below the desired interaction rate without violating the adiabatic criteria. As a consequence, there does not exist a regime where the gate can be performed. We solved this issue by proposing to shelve the qubit into a metastable ground state that is uncoupled to the cavity, and then we demonstrated that a large high-fidelity regime exists.

 \section{Results and Discussion} \label{sec:compare}
 \subsection{Scheme comparison}
 \begin{figure}
\centering
	\mbox{\hspace{-10mm}\includegraphics[width=7.5cm]{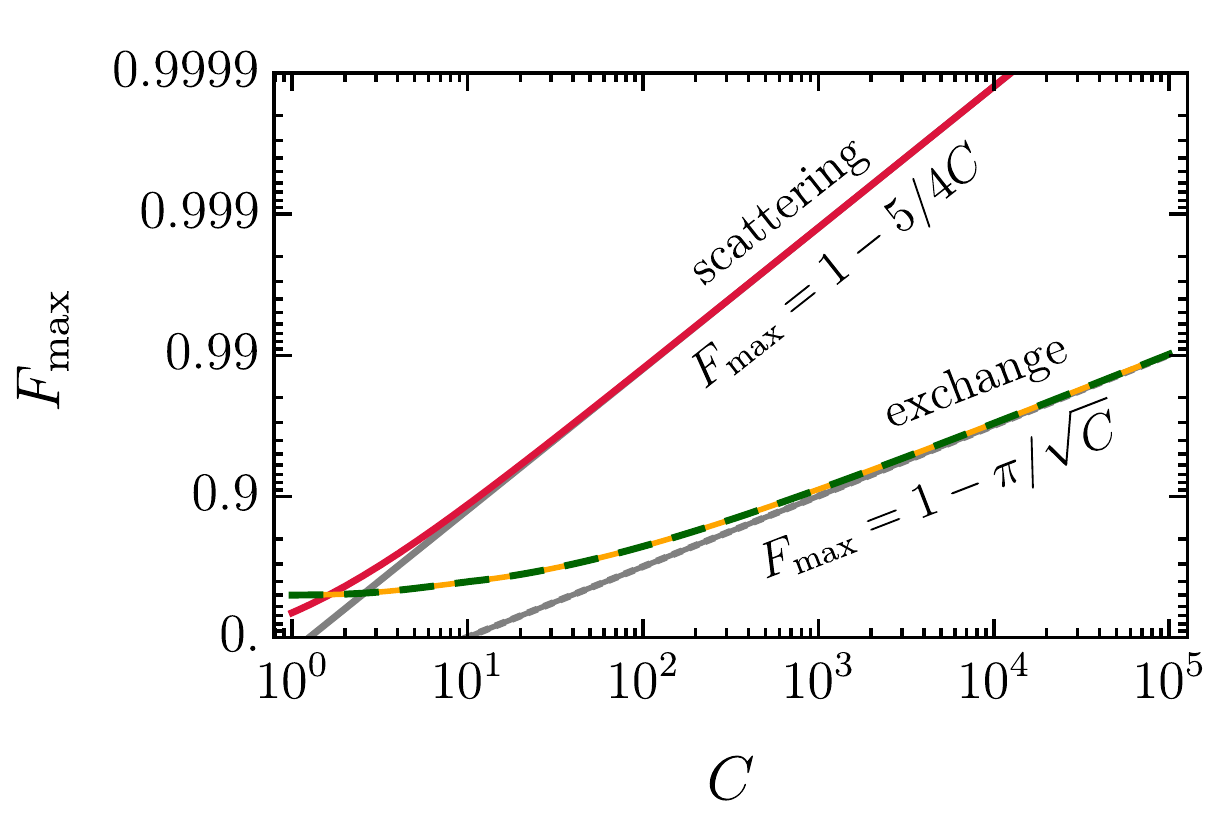}}
	\caption{Cooperativity-limited fidelity $F_\text{max}$ for phase gates based on photon scattering and virtual photon exchange. The simple photon exchange (orange line) and Raman photon exchange (dashed green line) have the same $F_\text{max}$ values. The straight gray lines represent the $F_\text{max}$ scaling in the limit of large $C$. These first-order approximations underestimate the maximum fidelity when the cooperativity is low, but still give a good estimate when the fidelity is larger than 0.8.}
	\label{fig:cooperativityscaling}
\end{figure}

The maximum fidelity scaling with cavity cooperativity is very different for the scattering scheme compared to the virtual photon exchange schemes (see Fig.~\ref{fig:cooperativityscaling}). In the photon scattering scheme the detection of the output photon heralds the gate and makes its fidelity independent of all sources of photon loss. Therefore, this scheme has the highest maximum fidelity with a scaling of $1-5/4C$. However, as a result of heralding, this scheme is probabilistic. On the other hand, the maximum fidelity of the virtual photon exchange schemes scales like $1-\pi/\sqrt{C}$,  but these schemes realize deterministic gates that do not rely on single-photon generation and detection.

\begin{figure}
\centering
     \begin{flushleft}\hspace{5mm}{\large(a)}\end{flushleft}\vspace{-5mm}
 	\mbox{\hspace{-8mm}\includegraphics[width=6.8cm]{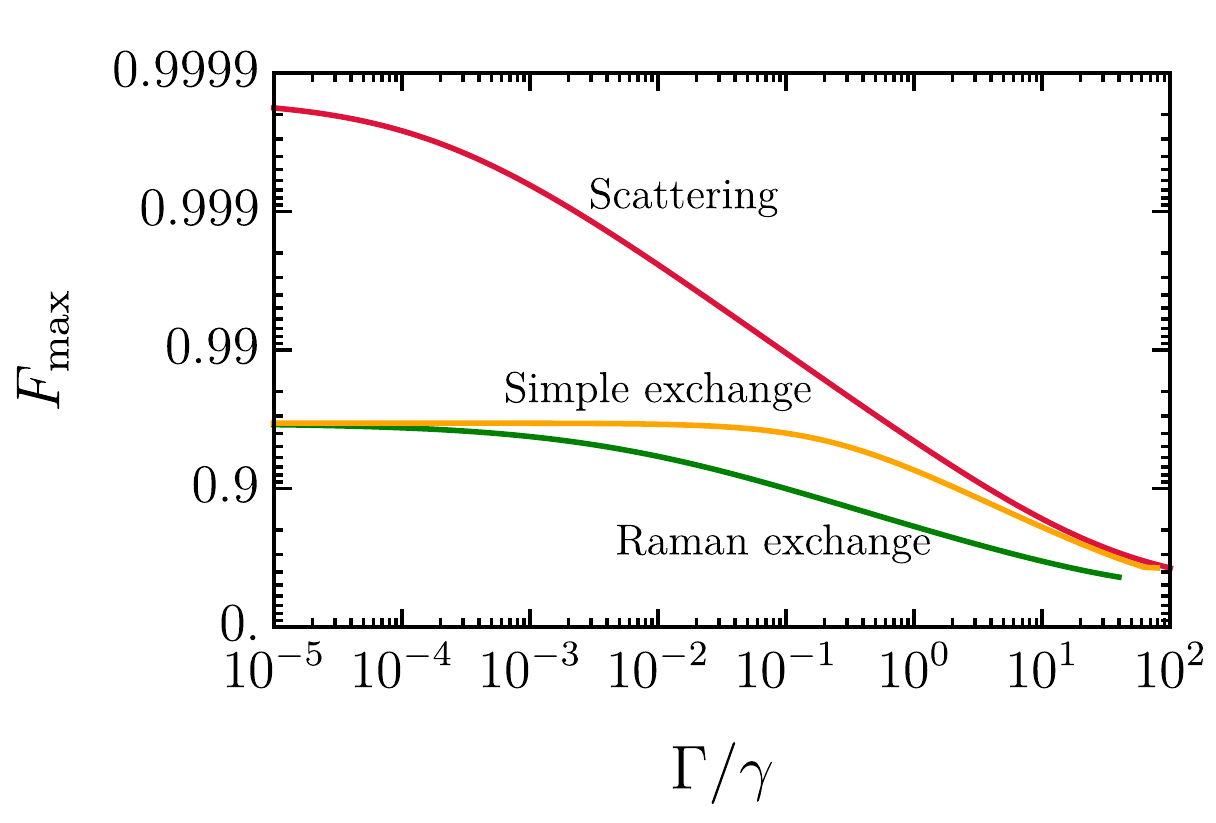}}\vspace{-5mm}
     \begin{flushleft}\hspace{5mm}{\large(b)}\end{flushleft}\vspace{-4mm}
 	\mbox{\hspace{-5mm}\includegraphics[width=6.5cm]{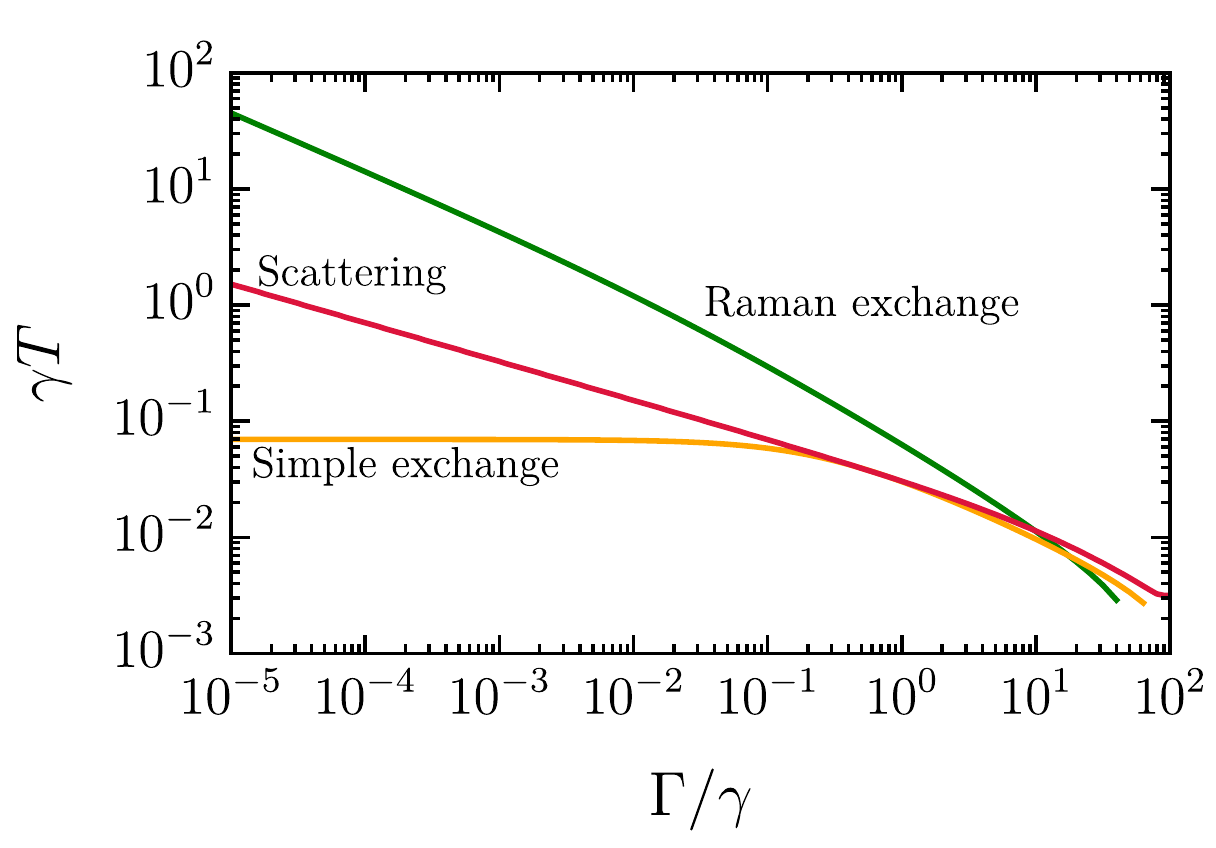}}
	\caption{(a) Maximum fidelity $F_\text{max}$ as a function of effective qubit decoherence rate $\Gamma$ for the scattering scheme (red curve), simple virtual photon exchange scheme (orange curve), and the Raman virtual photon exchange scheme (green curve) using a system in the bad-cavity regime $g/\kappa=10^{-1}$ with a cooperativity of $C=8000$. (b) Gate time corresponding to the maximum fidelity in panel (a).}
	\label{fig:exchangecompare}
\end{figure}

For each scheme, an increase in qubit decoherence will reduce the maximum attainable fidelity. To partially mitigate this effect, it is possible to reduce the total gate time. However, reducing the gate time below the optimal value will also reduce the fidelity. These two opposing effects create an intermediate optimal gate time that maximizes fidelity as a function of the effective decoherence rate of the qubits. This decoherence-limited maximum fidelity and corresponding optimal gate time have noticeably different trends for each scheme (see Fig.~\ref{fig:exchangecompare}). In the following, we discuss other pros and cons for each scheme in more detail.
 
\emph{Photon scattering.---} This scheme requires two nearly-identical quantum systems that must both have transitions resonant with the cavity. Having individual spectral control may require spatial resolution of the systems, which is a disadvantage for nanoscale devices. An advantage for this scheme is that the systems are not optically excited when performing the gate. Hence, this scheme could be of interest in systems with lower cavity cooperativity and some optical pure dephasing. Quantum dot devices are particularly suited to this scheme for the latter reasons, but also because a similar device could be used to efficiently generate the required single photons, providing a cohesive platform.
RE ions may also be promising candidates for this scheme. Single RE emitters have been observed \cite{zhong}, and when coupled to a high quality factor cavity the system could provide a cavity-cooperativity large enough to achieve fast controlled phase-flip gates with a high fidelity.

The probability of heralding will depend on the efficiency of available indistinguishable single-photon sources and detectors. On-demand sources with high photon indistinguishability and single-photon purity have been demonstrated \cite{source1, source3,ding2016demand}. In addition, highly efficient on-demand sources should become increasingly available with advances in deterministic fabrication \cite{source5}. The best commercially available sources provide an efficiency of around $10\%-30\%$ in practice, but these values are likely to improve in the near future \cite{wang2019towards}. Single-photon detector efficiency is also improving \cite{detector1, detector2}; superconducting single-photon nanowire detectors with efficiencies exceeding 90\% are becoming widely available. The overall success probability could be improved significantly if the photon source, detector, and cavity are all integrated on-chip \cite{integrated1, integrated2}. It is also possible to extend the scheme to perform non-local gates between multiple qubit-cavity systems (i.e., remote cavities). This ability can help with scalable quantum computing by naturally providing a connection between multiple qubits.

\emph{Simple virtual photon exchange.---} As with the scattering scheme, this scheme also requires the ability to tune the systems' optical transitions into resonance. This can be accomplished, for example, by using an AC Stark effect provided that systems are spatially resolved or by using a large electric or magnetic field gradient if they are not spatially separated. However, after tuning the systems, spatial resolution is still required to excite only one system to the excited state without affecting the other qubit. To avoid this requirement, it might be possible to excite one system before tuning them into resonance. As a result, the time it takes to tune the systems into resonance should be much faster than the phase evolution time of the system yet slow enough to remain adiabatic. Otherwise, the phase evolves while tuning the systems, which may limit the gate fidelity.


This scheme benefits from the exchange of virtual photons; therefore, the cavity induced relaxation can be avoided. However, a limiting factor of the scheme is still the excited state lifetime of the systems. To perform the gate, it is necessary for the excited system to remain excited for a time that is long enough compared to the gate time. Otherwise, the system decays before the phase-flip gate takes place. This effect is the primary cause of the reduced cooperativity-limited fidelity of $1-\pi/\sqrt{C}$ compared to the scattering scheme. On the other hand, the simple exchange scheme can be very fast, reducing the impact of qubit dephasing. This scheme is particularly suited to systems with little optical pure dephasing and small phonon sidebands, such as the group-IV defects in diamond \cite{group6defects} and rare-earth ions \cite{Er167, ultraslow}.

It is also possible to perform this scheme using a $\Lambda$ system. However, in the 4-level system that we have presented, tuning opposite transitions into resonance can prevent the requirement for the spatial resolution in systems with different polarization for opposite transitions, provided that both transitions can still be coupled to the same cavity mode.

\emph{Raman virtual photon exchange.---} The main advantage of this scheme is the ability to adjust the frequency and intensity of driving fields A and B to allow for a difference between the optical transition frequencies of the systems. As a downside, this method requires an additional metastable state to shelve one qubit. 
In addition, for a large $\kappa$, the optimal detuning $\delta$ must be quite large. This could be a major limitation for some systems with multiple close optical transitions, such as rare earth ions.

With the correct parameters, the Raman scheme can be performed while maintaining the spectral resolution of the system optical transitions. This is a huge advantage for solid-state microcavity systems where emitters are often quite different and their close proximity may not allow for spatial addressing. The main trade-off for this advantage is an increase in total gate time compared to the simple virtual photon exchange, making it more susceptible to decoherence of the metastable state.

The target qubit can either be shelved in a metastable state in the ground state or in a second uncoupled excited state. However, it is preferable to shelve the qubit in a ground state as the decoherence rate of the ground states are usually less than the excited states. A spin-triplet ground-state system with a relatively long spin-coherence time and good optical properties, such as in a neutrally-charged silicon-vacancy center in diamond \cite{SiV}, would provide the ideal structure for this scheme.

In systems where shelving is not feasible, one could establish a Raman coupling directly between the two quantum systems \cite{You}, rather than a Raman coupling for each of the qubits individually, as explained in our protocol. Such a scheme would require a weak external laser field ($\Omega < g$) and spatial-separated nearly-identical quantum systems.

\subsection{Comparison of all three gate schemes for $^{171}$Yb:YVO}
\begin{figure}
\includegraphics[width=5.5cm]{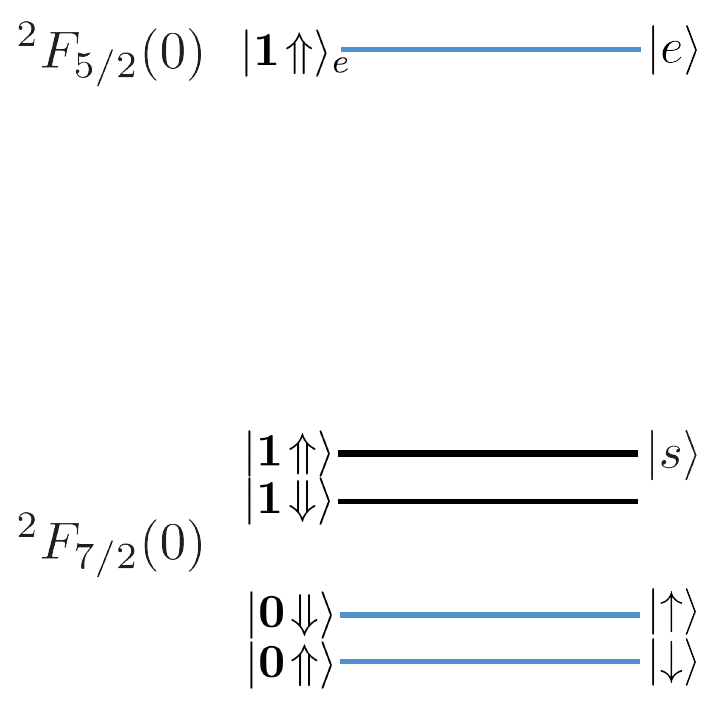}
\caption{Energy level structure of $^{171}$Yb:YVO in the presence of an external magnetic field along the c-axis of the crystal. Here, $\ket{\boldsymbol{0}}\equiv\ket{m_s=1/2}$ and $\ket{\boldsymbol{1}}\equiv\ket{m_s=-1/2}$ are electron spin states and  $\ket{\Uparrow}\equiv\ket{m_I=1/2}$ and $\ket{\Downarrow}\equiv\ket{m_I=-1/2}$ are nuclear spin states.
We use the $\ket{\boldsymbol{0}\Downarrow}$,  $\ket{\boldsymbol{0}\Uparrow}$ and $\ket{\boldsymbol{1}\Uparrow}$ hyperfine states as the qubit states $\ket{\uparrow}$ and $\ket{\downarrow}$ and the shelving state $\ket{s}$ respectively.}\label{fig:Yb}
\end{figure}

We consider $^{171}$Yb ions doped into a yttrium orthovanadate (YVO) crystal as an example system to compare the three different gates. The energy level structure of this ion in the presence of an external magnetic field is shown in Fig.~\ref{fig:Yb}. Note that, the figure only shows the lowest excited state level. Here, we refer to the two lowest ground state hyperfine levels as the qubit states $\ket{\downarrow}$ and $\ket{\uparrow}$.

For an ensemble of $^{171}$Yb ions in YVO, the excited state decay rate is $\gamma=2\pi\times596$ Hz \cite{Yb}. In addition, the spin coherence time of $T_2=6.6$~ms has been measured (for $B=440$~mT ) \cite{Yb}. For a single Yb coupled to a YVO photonic crystal cavity, it has been shown that the spin coherence time can be further increased to 30~ms using a Carr-Purcell-Meiboom-Gill (CPMG) decoupling sequence \cite{Yb-30}.
	
In the following, we estimate the maximum gate fidelity and the corresponding gate time for each scheme when assuming $g/\kappa=10^{-1}$ and  $C=50~000$.
	
\emph{Photon scattering.---} Considering $\Gamma=1/(2T_2)$, where $T_2=6.6$ ms, $T=1/\gamma$, $\delta_p= 30\gamma$ and $\delta_{\epsilon_A}=\delta_{\epsilon_B}=0$,  the fidelity of the photon reflection scheme is $F_\text{max}=0.98$ and the gate time is $T=267~\mu$s.
	
\emph{Simple virtual photon exchange.---}
As discussed previously in Sec.~\ref{sec:compare}, it is also possible to perform the simple virtual photon exchange scheme using a $\Lambda$ type system. Here we consider a three level system and, to perform the gate we bring the $\ket{\uparrow}-\ket{e}$ transitions of the two ions into resonance with each other (instead of tuning opposite transitions). In this case, considering $\delta_{eg}=0.2$ GHz \cite{Yb}, $\Gamma=1/(2T_2)+\gamma^\star/2$ (here $\gamma^\star=9$ KHz is calculated by the relation $\gamma^\star=1/T_{2,O}-\gamma/2$
where $T_{2,O}=91~\mu$s is the optical coherence time for $B=500$ mT \cite{Yb}), and $\Delta_{\epsilon}=0$ we get $F_\text{max}=0.952$ and $T=7.5~\mu$s.
	
\emph{Raman virtual photon exchange.---}
For this scheme, we consider the $\ket{\boldsymbol{1}\Uparrow}$ hyperfine ground state as the shelving level as shown in Fig.~\ref{fig:Yb}. Assuming $\Gamma=1/(2T_2)$, $\delta_\epsilon=0$,  $\Delta_A=\Delta_B$ and $\Omega=0.1\Delta$, the optimal fidelity and the gate time are $F_\text{max}=0.93$ and $T=750~\mu$s, respectively.

Increasing the cavity cooperativity will increase the fidelity and decrease the gate time of the simple and Raman virtual photon exchange schemes further. On the other hand, to improve the fidelity and the gate time of the photon scattering scheme, a photon with a smaller FWHM duration is required. 
Although the properties and level structure of $^{171}$Yb:YVO allows performing all three schemes, the most suitable scheme for this system should be selected according to the gate requirements and experimental restrictions. As an example, if the optical transition frequencies of Yb ions are unequal, one should perform the gate using the Raman scheme. The simple virtual exchange scheme, on the other hand, is the best option to perform a fast gate. And finally, the photon scattering scheme can lead to a probabilistic but high fidelity gate.

\section{Conclusion}\label{sec:conclusion}
Using the cavity-assisted interactions, we proposed and compared three schemes to perform controllable phase-flip gates between two qubits.
The first scheme works better for systems with an integrated design and when performing a high fidelity gate is more important than a deterministic gate. If one looks for a deterministic gate, however, either the simple or Raman virtual photon exchange schemes should be considered. In cases where the quantum systems are not resonant, the Raman exchange is the best scheme. On the other hand, the simple exchange can be suitable for systems with more severe qubit dephasing but little pure dephasing of the optical transition.

Looking forward, our promising results on the photon scattering gate may provide further motivation for integrating sources and cavities on chip. Moreover, the fidelity of the simple and Raman virtual photon exchange schemes could be improved using the quantum Zeno effect \cite{zeno}. In this technique, by observing possible photons emitted by the cavity at frequent time intervals using an efficient single-photon detector, the system can be forced to follow the adiabatic evolution \cite{pachos}. Detecting a leaked photon also indicates a failed gate and improves fidelity.

Developing quantum information processing nodes using cavity-mediated gates is an important step towards the implementation of quantum networks. By outlining the benefits and limitations of different approaches to this goal, we provided a framework to identify and tailor two-qubit gate schemes for a given system. This will accelerate the development of platforms that could form the basis for a future quantum internet.
 
\section*{Acknowledgments}
The authors thank I. Craiciu, A. Faraon, S. Goswami, H. Ollivier, and T. Zhong for helpful discussions.
This work was supported by the Natural Sciences and Engineering Research Council of Canada (NSERC) through its Discovery Grant,  Canadian Graduate Scholarships and CREATE programs, and by Alberta Innovates Technology Futures (AITF).

\section*{COMPETING INTERESTS}
The authors declare no competing interests.

\setcounter{equation}{0}
\renewcommand{\theequation}{A\arabic{equation}}
\renewcommand{\appendixname}{APPENDIX}
\appendix
\section{APPROACH}\label{Approach}
\subsection{Fidelity}
For each scheme, we define two initially independent quantum systems that each include qubit states denoted $\ket{\uparrow}$ and $\ket{\downarrow}$. The total two-qubit space is spanned by the four cannonical product space states $\ket{\uparrow\uparrow}$, $\ket{\uparrow\downarrow}$, $\ket{\downarrow\uparrow}$, and $\ket{\downarrow\downarrow}$. In our analysis, we define the gate fidelity $F_\text{gate}$ to be the fidelity after applying the gate operations to the initial product state $\ket{\psi(0)}=(1/2)(\ket{\uparrow}+\ket{\downarrow})_A\otimes(\ket{\uparrow}+\ket{\downarrow})_B$:
\begin{equation}
\label{fidelityDef}
    F_\text{gate} = \sqrt{\bra{\psi(T)}\hat{\rho}(T)\ket{\psi(T)}},
\end{equation}
where $\rho(T)$ (or $\ket{\phi(T)}\bra{\phi(T)}$ in the case of a pure state) is the imperfect final state and $\ket{\psi(T)}$ is the expected ideal pure state after applying the gate operation with duration time $T$. For example, if the gate operation takes $\ket{\downarrow\downarrow}\longrightarrow-\ket{\downarrow\downarrow}$ relative to the remaining three product states, the ideal state is $\ket{\psi(T)}=(1/2)(\ket{\uparrow\uparrow}+\ket{\uparrow\downarrow}+\ket{\downarrow\uparrow}-\ket{\downarrow\downarrow})$. This choice of initial state serves to represent an average gate fidelity because it takes into account the effect of the gate on each product state amplitude in addition to the relative phases between them. It also represents the fidelity expected when using the gate to generate maximally entangled states. However, certain initial states may result in higher or lower fidelity than this definition predicts. For example, the initial state $\ket{\uparrow}_A\otimes(\ket{\uparrow}+\ket{\downarrow})_B/\sqrt{2}$ for the above example could have above-average fidelity since it will not experience infidelity due to the imperfect phase-flip operation on $\ket{\downarrow\downarrow}$. On the other hand, $\ket{\downarrow}_A\otimes(\ket{\uparrow}+\ket{\downarrow})_B/\sqrt{2}$ could experience below-average fidelity due to the absence of contribution from the less-stringent evolution on $\ket{\uparrow\uparrow}$ and $\ket{\uparrow\downarrow}$.

\subsection{Decoherence}
To simplify the analysis and to focus on the intrinsic high-performance limitations for each scheme, we assume that decoherence processes other than cavity dissipation and spontaneous emission occur on a timescale much longer than the gate time. These additional processes include qubit decoherence and possibly pure dephasing of the optical transition. To capture these small effects, we describe the effect of any of these additional processes by a single effective decoherence rate $\Gamma$. The exact form of $\Gamma$ may be different depending on the scheme and on the dominant source of additional decoherence for a system operating under a given scheme. For example, regardless of the scheme, the effective qubit decoherence rate $\Gamma$ must be at least limited by the qubit relaxation rate $\gamma_{\uparrow\downarrow}$ and pure dephasing rate $\gamma_{\uparrow\downarrow}^\star$. Consider the following decoherence master equation:
\begin{equation}
    \dot{\rho} = \gamma_{\uparrow\downarrow}\mathcal{D}(\hat{\sigma})\rho(t)+2\gamma_{\uparrow\downarrow}^\star\mathcal{D}(\hat{\sigma}^\dagger\hat{\sigma})\rho(t),
\end{equation}
where $\mathcal{D}(\hat{\sigma})\hat{\rho}=\hat{\sigma}\hat{\rho}\hat{\sigma}^\dagger-\{\hat{\sigma}^\dagger\hat{\sigma},\hat{\rho}\}/2$ for $\hat{\sigma}\ket{\uparrow}=\ket{\downarrow}$. If we wish to maintain the coherence of an initial state $\ket{\psi(0)}=(\ket{\uparrow}+\ket{\downarrow})/\sqrt{2}=\ket{\psi(T)}$, the fidelity of the final state $\hat{\rho}(T)$ will be
\begin{equation}
    \sqrt{\bra{\psi(T)}\hat{\rho}(T)\ket{\psi(T)}} \simeq 1 - \Gamma T,
\end{equation}
when expanding to first order in $\gamma_{\uparrow\downarrow}T\ll 1$ and $\gamma_{\uparrow\downarrow}^\star T\ll 1$ where $\Gamma = \gamma_{\uparrow\downarrow}/8+\gamma_{\uparrow\downarrow}^\star/4$. In most real applications, the effective decoherence rate $\Gamma$ will be dominated by the largest source of additional decoherence for that particular system or scheme-system combination.

\subsection{Non-Hermitian dynamics}

In the virtual photon exchange schemes, we take advantage of non-Hermitian Hamiltonians to include cavity dissipation and spontaneous emission as opposed to solving the full master equation. This allows us to capture the effects of finite cavity cooperativity while still allowing for simple and accurate analytically tractable solutions.

Dynamics from non-Hermitian Hamiltonians cannot capture an increase in population of the ground state due to a decay event. To illustrate this, consider the Master equation
\begin{equation}
    \dot{\rho} = -i[\hat{H},\hat{\rho}] + \gamma\mathcal{D}(\hat{\sigma})\hat{\rho} + \kappa\mathcal{D}(\hat{a})\hat{\rho},
\end{equation}
where we take $\hbar=1$. This can be rewritten as \cite{daley2014quantum}:
\begin{equation}
\begin{aligned}
    \dot{\rho} &= -i\left[\hat{H},\hat{\rho}\right] - \frac{1}{2}\{\gamma \hat{\sigma}^\dagger\hat{\sigma}+\kappa\hat{a}^\dagger\hat{a},\hat{\rho}\} + \gamma\hat{\sigma}\hat{\rho}\hat{\sigma}^\dagger + \kappa\hat{a}\hat{\rho}\hat{a}^\dagger\\
    &=-i\left(\hat{\mathcal{H}}_\text{eff}\hat{\rho}-\hat{\rho}\hat{\mathcal{H}}_\text{eff}^\dagger\right)+\gamma\hat{\sigma}\hat{\rho}\hat{\sigma}^\dagger + \kappa\hat{a}\hat{\rho}\hat{a}^\dagger,
\end{aligned}
\end{equation}
where
\begin{equation}
    \hat{\mathcal{H}}_\text{eff} = \hat{H} -\frac{i}{2}\left(\gamma \hat{\sigma}^\dagger\hat{\sigma}+\kappa\hat{a}^\dagger\hat{a}\right)
\end{equation}
is the effective non-Hermitian Hamiltonian that describes the amplitude decay of $\hat{\sigma}$ and $\hat{a}$.

The solution $\ket{\phi(t)}$ under the effective Hamiltonian is the unnormalized pure state trajectory for a successful gate and this trajectory occurs with probability $p=\braket{\phi(t)|\phi(t)}$. The terms $\gamma\hat{\sigma}\hat{\rho}\hat{\sigma}^\dagger$ and $\kappa\hat{a}\hat{\rho}\hat{a}^\dagger$ in the master equation cause a recycling of population into the ground state after a decay event. Thus the total master equation solution is given by $\hat{\rho}(t)=\ket{\phi(t)}\!\bra{\phi(t)}+(1-p)\hat{\rho}_{\gamma\kappa}(t)$ where $\hat{\rho}_{\gamma\kappa}(t)$ is the state of the system at time $t$ given that at least one emission event occurred. The final fidelity after completing the gate of duration $T$ is then
\begin{equation}
    F_\text{gate} = \sqrt{p F_0^2 + (1-p)F_{\gamma\kappa}^2}
\end{equation}
where $F_0=|\braket{\phi(T)|\psi(T)}|/\sqrt{p}$ is the fidelity after a successful gate and $F_{\gamma\kappa}=\sqrt{\bra{\psi(T)}\hat{\rho}_{\gamma\kappa}(T)\ket{\psi(T)}}$ is the potentially non-zero fidelity after a failed gate.

By only solving the effective non-Hermitian Hamiltonian part of the master equation, we make the approximation that $F_\text{gate}\simeq \sqrt{p} F_0$. This approximation is accurate when $p\simeq 1$ and hence when $\sqrt{p}F_0\simeq 1$. The precision of this approximation depends on $F_0$ and $F_{\gamma\kappa}$ for a given implementation. Since $F_{\gamma\kappa}< F_0$ for the schemes we study, this approximation is also valid to explore the cooperativity scaling of the fidelity limits. We comment on the accuracy of this approximation for the specific cases of the virtual photon exchange schemes in the following appendix section.

\setcounter{equation}{0}
\renewcommand{\theequation}{B\arabic{equation}}
\renewcommand{\appendixname}{APPENDIX}
\section{FIDELITY CALCULATIONS}\label{AppendixB}
\setcounter{subsection}{0}
\subsection{Photon scattering}

In this scheme, the probability that an incident photon excites either qubit is low. Therefore, the quantum Langevin equations for the photon (quantum system) excitation amplitude(s) $a(t)$ ($s_k(t)$, $k\in \{A,B\}$) can be written as \cite{Obrien,giant, quantum}  
\begin{equation}
\begin{aligned}
\dot{a}(t)&=-\frac{\kappa}{2}a(t)+g_A s_A(t)+g_B s_B(t)-\sqrt{\kappa}a_\text{in}(t),\\
\dot{s}_A(t)&=-g_A a(t)+\left(-\frac{\gamma}{2}-i\Delta_A\right)s_A(t),\\
\dot{s}_B(t)&=-g_Ba(t)+\left(-\frac{\gamma}{2}-i\Delta_B\right)s_B(t),
\end{aligned}
\end{equation}
where $a_\text{in}(t)$ is the input photon field, $g_k$ is the cavity coupling rate for the $k^\text{th}$ quantum system, and $\Delta_k$ is the detuning between the $\ket{\downarrow}\rightarrow\ket{e}$ transition of the $k^\text{th}$ quantum system and the bare cavity mode. Using the input-output relation $a_\text{out}=\sqrt{\kappa}a+a_\text{in}$, the ratio of output and input field for a plane wave input is
\begin{equation}
\frac{a_\text{out}(\omega)}{a_\text{in}(\omega)}= 1-\frac{\kappa}{\kappa/2+g^2_A/r_A+g^2_B/r_B-i\omega},
\end{equation}
where $r_k=\gamma/2+i(\Delta_k-\omega)$ and $\omega$ is the plane wave frequency detuning from the cavity resonance. This expression is valid in both the strong-coupling and bad-cavity regimes \cite{Obrien}. Using the above general expression for the photon amplitude of plane wave $\ket{\omega}$, we can write the amplitude $s_{ij}(\omega)$ (where $i,j\in\{\uparrow,\downarrow\}$) expected for each initial qubit product state $\ket{\uparrow\uparrow}$, $\ket{\uparrow\downarrow}$, $\ket{\downarrow\uparrow}$, and $\ket{\downarrow\downarrow}$ so that $\ket{ij}\ket{\omega}\longrightarrow s_{ij}(\omega)\ket{ij}\ket{\omega}$. This can be done by setting the $\Delta_k$ zero for $\ket{\uparrow}_k$ and non-zero but large for $\ket{\downarrow}_k$. Under the assumptions that $g_A= g_B= g$ and $\Delta_A$, $\Delta_B \gg \kappa$ (when nonzero) we have
\begin{equation}
\label{eq_ratio}
\begin{aligned}
s_{\uparrow\uparrow}(\omega)&=\left.\frac{a_\text{out}(\omega)}{a_\text{in}(\omega)}\right|_{\Delta_A=0,\Delta_B=0}\\&=1-\frac{2\kappa(\gamma-2i\omega)}{2\kappa\gamma C+(\kappa-2i\omega)(\gamma-2i\omega)},\\
s_{\uparrow\downarrow}(\omega)&=\lim_{\Delta_B\rightarrow\infty}\left.\frac{a_\text{out}(\omega)}{a_\text{in}(\omega)}\right|_{\Delta_A=0} \\
    &= 1-\frac{2\kappa(\gamma-2i\omega)}{\kappa\gamma C+(\kappa-2i\omega)(\gamma-2i\omega)},\\
s_{\downarrow\uparrow}(\omega)&=\lim_{\Delta_A\rightarrow\infty}\left.\frac{a_\text{out}(\omega)}{a_\text{in}(\omega)}\right|_{\Delta_B=0}\\
    &= 1-\frac{2\kappa(\gamma-2i\omega)}{\kappa\gamma C+(\kappa-2i\omega)(\gamma-2i\omega)},\\
s_{\downarrow\downarrow}(\omega)&=\lim_{\Delta_B\rightarrow\infty,\Delta_A\rightarrow\infty}\frac{a_\text{out}(\omega)}{a_\text{in}(\omega)} = -1-\frac{4i\omega}{\kappa-2i\omega}.
\end{aligned}
\end{equation}
where $C=4g^2/\kappa\gamma$. Although we do not present it here, the above set of equations could include finite quantum system detunings for systems $A$ and $B$ by evaluating the ratio $a_\text{out}/a_\text{in}$ for $\Delta_A=\delta_{\epsilon_A}$ and $\Delta_B=\delta_{\epsilon_B}$ instead of $\Delta_A=0$ and $\Delta_B=0$, where appropriate. To illustrate how these amplitudes indicate a controlled phase gate, consider the ideal case where we have a perfect plane wave exactly resonant with the cavity so that $\omega=0$. Then the amplitudes reduce to
\begin{equation}
\label{eq_ratio}
\begin{aligned}
s_{\uparrow\uparrow}(0)&= 1-\frac{2}{2C+1},\\
s_{\uparrow\downarrow}(0)&= 1-\frac{2}{C+1},\\
s_{\downarrow\uparrow}(0)&= 1-\frac{2}{C+1},\\
s_{\downarrow\downarrow}(0)&= -1.
\end{aligned}
\end{equation}
From this it is clear that when $C\gg1$ these ratios converge to 1, 1, 1 and -1, respectively.

In reality, some deviation from the ideal conditions are expected. In particular, we consider a finite Gaussian bandwidth photon with a standard deviation $\sigma_p$ and a possible small cavity resonance error of $\delta_p$. Even though the final spin state is pure for a plane wave, the spin-frequency entanglement captured by the frequency-dependent amplitudes $s_{ij}(\omega)$ causes some reflection-induced spin dephasing. To correct for a finite bandwidth photon, we consider an initial photon state $\ket{p}=\int d\omega f(\omega)\ket{\omega}$ where
\begin{equation}
    |f(\omega)|^2 = \frac{1}{\sigma_p\sqrt{2\pi}}e^{-(\omega-\delta_p)^2/2\sigma_p^2}.
\end{equation}
For an initial spin state $(1/2)(\ket{\uparrow}+\ket{\downarrow})_A\otimes(\ket{\uparrow}+\ket{\downarrow})_B$, the joint spin-photon state after reflection is
\begin{equation}
    \ket{\phi(T)}_\text{sp}=\frac{1}{2}\int d\omega \sum_{ij}s_{ij}(\omega)f(\omega)\ket{ij}\ket{\omega}.
\end{equation}
where we take the total gate time $T$ to be twice the FWHM of the photon duration: $T=8\pi\sqrt{2\ln{2}}/\sigma_p$. The reduced spin density matrix $\hat{\rho}$ can then be obtained by tracing out the state of the photon $\hat{\rho}(T) = \text{Tr}_p\left(\ket{\phi(T)}\!\bra{\phi(T)}_\text{sp}\right)$. This gives
\begin{equation}
\label{reducedmatrix}
    \hat{\rho}(T) = \frac{1}{4}\int d\omega \sum_{ij}\sum_{kl}s_{ij}(\omega)s_{kl}^*(\omega)|f(\omega)|^2\ket{ij}\bra{kl}.
\end{equation}


After a single photon reflects off the cavity, the final state of the two-qubit system can be compared with the ideal state $\ket{\psi(T)}=(1/2)(\ket{\uparrow\uparrow}+\ket{\uparrow\downarrow}+\ket{\downarrow\uparrow}-\ket{\downarrow\downarrow})$ to give the total gate fidelity $F_\text{gate}$ from equation (\ref{fidelityDef}). By following this procedure using (\ref{reducedmatrix}) to take into account small imperfections due to nonzero $\delta_p$, $\sigma_p$, and $\delta_{\epsilon_k}$, we derived the total gate fidelity $F_\text{gate}$ as presented in equation (1) of the main text. This was done analytically by first expanding the amplitudes $s_{ij}$ in terms of small $\omega/C\gamma$ and then integrating over the Gaussian photon profile.

We also note that, in the limit that $\sigma_p$ and $\delta_p$ are small, the amplitudes from equation (\ref{eq_ratio}) immediately give the cooperativity-limited maximum fidelity of
\begin{equation}
    F_\text{max} = 1 - \frac{1}{C+1} - \frac{1}{4 C+2}\simeq 1- \frac{5}{4C}.
\end{equation}

\subsection{Simple virtual photon exchange}
For this scheme, we begin with two four-level systems coupled to a single cavity mode. The general Hamiltonian that governs the evolution is $\hat{H} = \hat{H}_A + \hat{H}_B+\hat{H}_C + \hat{H}_I$. The quantum system $\hat{H}_k$ is given by
\begin{equation}
    \hat{H}_k = \omega_k\hat{\sigma}_{\uparrow_k}^\dagger\hat{\sigma}_{\uparrow_k}+(\omega_k-\omega_e)\hat{\sigma}_{\downarrow_k}^\dagger\hat{\sigma}_{\downarrow_k}-\omega_g\hat{\sigma}_{\uparrow\downarrow_k}^\dagger\hat{\sigma}_{\uparrow\downarrow_k},
\end{equation}
where $\omega_k$ is the frequency separation between $\ket{\uparrow}_k$ and $\ket{e_2}_k$, $\omega_e$ is the separation between $\ket{e_1}_k$ and $\ket{e_2}_k$, and $\omega_g$ is the separation between $\ket{\uparrow}_k$ and $\ket{\downarrow}_k$. Also, $\hat{\sigma}_{\downarrow_k}\ket{e_1}_k=\ket{\downarrow}_k$, $\hat{\sigma}_{\uparrow_k}\ket{e_2}=\ket{\uparrow}_k$, and $\hat{\sigma}_{\uparrow\downarrow_k}\ket{\downarrow}_k=\ket{\uparrow}_k$ (see figure 3 of the main text). The cavity homogeneous evolution is $\hat{H}_C = \omega_C\hat{a}^\dagger\hat{a}$ for cavity frequency $\omega_C$, cavity photon anihilation (creation) operator $\hat{a}$ ($\hat{a}^\dagger$), and the interaction part is given by
\begin{equation}
    \hat{H}_I = \sum_{j\in{\uparrow,\downarrow}}\sum_{k\in{A,B}}g_{j_k}\hat{\sigma}_{j_k}^\dagger\hat{a} + \text{h.c.},
\end{equation}
where $g_{\downarrow_k}$ is the cavity coupling rate of the $\ket{\downarrow}\rightarrow\ket{e_1}$ transition to the cavity mode and $g_{\uparrow_k}$ is the cavity coupling rate of the $\ket{\uparrow}\rightarrow\ket{e_2}$ transition to the cavity mode. The dissipation is governed by the Lindblad master equation
\begin{equation}
    \dot{\rho} = -i[\hat{H},\hat{\rho}] +\kappa\mathcal{D}(\hat{a})\hat{\rho}+ \sum_{k,j}\gamma_{j_k}\mathcal{D}(\hat{\sigma}_{j_k})\hat{\rho}
\end{equation}
where $\mathcal{D}(\hat{A})\hat{\rho}=\hat{A}\hat{\rho}\hat{A}^\dagger -\{\hat{A}^\dagger\hat{A},\hat{\rho}\}/2$, $\kappa$ is the decay rate of the cavity photon, $\gamma_{j_k}$ are the decay rates of the $\ket{e_1}_k\rightarrow\ket{\downarrow}_k$ and $\ket{e_2}_k\rightarrow\ket{\uparrow}_k$ transitions. In the following, we assume $\gamma_{j_k} = \gamma$ for all $j$ and $k$. The corresponding effective non-Hermitian Hamiltonian is then
\begin{equation}
    \hat{\mathcal{H}}_\text{eff} = \hat{H} - \frac{i}{2}\left[\kappa\hat{a}^\dagger\hat{a}+ \gamma\sum_{k,j}\hat{\sigma}_{j_k}^\dagger \hat{\sigma}_{j_k}\right]
\end{equation}
Note that the effective Hamiltonian for dissipation due to $\gamma$ does not discriminate between events that emit into $\ket{\uparrow}$ or $\ket{\downarrow}$ because the recycling term is neglected. That is, $\gamma$ here represents the total decay rate of the excited states.

The total cavity-qubit system can be broken into four subsystems defined by the four basis states of the electronic ground states: $\{\ket{\uparrow\uparrow}, \ket{\downarrow\uparrow}, \ket{\uparrow\downarrow}, \ket{\downarrow\downarrow}\}$. To perform a control phase gate using a virtual photon interaction, quantum system A is excited at $\omega_A$ so that $\ket{\uparrow}_A\rightarrow\ket{e_2}_A$. This implies that $\ket{\uparrow\downarrow}\rightarrow\ket{e_2\downarrow}$ and $\ket{\uparrow\uparrow}\rightarrow\ket{e_2\uparrow}$. Hence we are concerned with the relative evolution within the two subsystems governed by $H_{\uparrow\downarrow}$ and $H_{\uparrow\uparrow}$. In our analysis, we assume that infidelity due to the fast excitation process $\ket{\uparrow}_A\rightarrow\ket{e_2}_A$ is much smaller than the infidelity due to the slower adiabatic virtual photon exchange process; we focus only on the phase rotation component of the protocol.

In the single-excited $\uparrow\downarrow$ subspace with the basis $\{\ket{e_2\downarrow 0},\ket{\uparrow\downarrow 1},\ket{\uparrow e_1 0}\}$, $H_{\uparrow\downarrow}$ can be written as
\begin{equation}
\label{hamupdown}
    \hat{H}_{\uparrow\downarrow} = \begin{pmatrix}
    0 & g_{\uparrow_A} & 0\\
    g_{\uparrow_A} &\Delta_A & g_{\downarrow_B}\\
    0 & g_{\downarrow_B} & \Delta_A-\Delta_B-\delta_{eg}\\    
    \end{pmatrix},
\end{equation}
where $\Delta_k = \omega_C - \omega_k$
and $\delta_{eg} = \omega_e-\omega_g$. In the single-excited $\uparrow\uparrow$ subspace with the basis $\{\ket{e_2\uparrow 0},\ket{\uparrow\uparrow 1},\ket{\uparrow e_2 0}\}$, $H_{\uparrow\uparrow}$ can be written as
\begin{equation}
\label{hamupdown}
    \hat{H}_{\uparrow\uparrow} = \begin{pmatrix}
    0 & g_{\uparrow_A} & 0\\
    g_{\uparrow_A} &\Delta_A & g_{\uparrow_B}\\
    0 & g_{\uparrow_B} & \Delta_A-\Delta_B\\    
    \end{pmatrix}.
\end{equation}
The last index of each combined-system state indicates the photon number in the cavity mode.

The evolution of the remaining subsystems is $H_{\downarrow\uparrow}=H_{\downarrow\downarrow}=0$ for the unexcited states $\ket{\downarrow\uparrow}$ and $\ket{\downarrow\downarrow}$. Note that since only two of the four basis states are evolving in this scheme; we are concerned with only the relative phase between $\ket{\uparrow\downarrow}$ and $\ket{\uparrow\uparrow}$. This is because any global phase for $\ket{\uparrow\downarrow}$ and $\ket{\uparrow\uparrow}$ can be eliminated by moving qubit $A$ into the correct rotating frame. Knowing this, we can simplify the total gate fidelity to \hbox{$F_\text{gate} = (F_\pi+1)/2$} where $F_\pi=|\bra{\phi(T)}(\ket{\uparrow\uparrow}-\ket{\uparrow\downarrow})|/\sqrt{2}$ is the fidelity of the relative $\pi$-phase gained between state $\ket{\uparrow\uparrow}$ and $\ket{\uparrow\downarrow}$ for the initial state $\ket{\psi(0)}=(\ket{\uparrow\uparrow}+\ket{\uparrow\downarrow})/\sqrt{2}$.

In the regime where $\Delta_k$ are much larger than the cavity coupling rates, the Hamiltonian $H_{\uparrow\uparrow}$ performs a $\pi$-phase rotation on $\ket{e_2\uparrow 0}$ if $\Delta_A-\Delta_B\simeq 0$. Alternatively, if $\Delta_A-\Delta_B\simeq\delta_{eg}$, $H_{\uparrow\downarrow}$ performs the $\pi$-phase on $\ket{e_2\downarrow 0}$. These two scenarios are equivalent and so, without loss of generality, we focus only on the case where the opposite spin transitions are resonant $\Delta_A-\Delta_B\simeq\delta_{eg}$. 

Since the cavity coupling rates may not be equal, it may be necessary to tune $\Delta_B$ to offset the different Stark shifts induced on each qubit by the cavity. By adiabatically eliminating the amplitude of state $\ket{\uparrow\downarrow 1}$, the optimal tuning of the unexcited qubit is found to be
\begin{equation}
\label{tuning}
    \Delta_B = \Delta + \frac{g_{\uparrow_A}^2-g_{\downarrow_B}^2}{\Delta}-\delta_{eg},
\end{equation}
where we write $\Delta_A=\Delta$ for simplicity. The corresponding excitation time required to achieve a $\pi$ phase is given by
\begin{equation}
    T = \pi\frac{\Delta}{g_{\uparrow_A}g_{\downarrow_B}}.
\end{equation}

High phase gate fidelity for virtual photon exchange is dependent on satisfying four main conditions: (1) the cavity detuning $\Delta$ must exceed the decay rate $\kappa$ of the cavity, (2) the gate time $T$ must not exceed the lifetime of the system excited state $1/\gamma$, and (3) the system should not be far into the strong-coupling regime $g/\kappa \leq 1$. Finally, (4) high fidelity operation depends critically on achieving the one-photon resonance condition. To capture how sensitive the fidelity is to errors in matching the resonance condition in equation (\ref{tuning}), we assume that $\Delta_B$ deviates from the ideal condition by some small value $\Delta_\epsilon$. That is, $\Delta_\epsilon = \Delta_B - \Delta - (g_{\uparrow_A}^2-g_{\downarrow_B}^2)/\Delta+\delta_{eg}$.

The effective non-Hermitian Hamiltonians corresponding to $\hat{H}_{\uparrow\downarrow}$ and $\hat{H}_{\uparrow\uparrow}$ are
\begin{equation}
\begin{aligned}
\hat{\mathcal{H}}_{\uparrow\downarrow}\!&=\!\hat{H}_{\uparrow\downarrow}\\&-\frac{i}{2}\left(\gamma\ket{e_2\!\downarrow \!0}\!\bra{e_2\!\downarrow\!0}\!+\!\gamma\ket{\uparrow\!e_1 0}\!\bra{\uparrow\!e_1 0}\!+\!\kappa\ket{\uparrow\downarrow \!1}\!\bra{\uparrow\downarrow\!1}\right)\\
\hat{\mathcal{H}}_{\uparrow\uparrow}\!&=\!\hat{H}_{\uparrow\uparrow}\\&-\frac{i}{2}\left(\gamma\ket{e_2\!\uparrow \!0}\!\bra{e_2\!\uparrow\!0}\!+\!\gamma\ket{\uparrow\!e_1 0}\!\bra{\uparrow\!e_1 0}\!+\!\kappa\ket{\uparrow\uparrow \!1}\!\bra{\uparrow\uparrow\!1}\right).\\
\end{aligned}
\end{equation}
By performing adiabatic elimination on the amplitude of $\ket{\uparrow\downarrow\!1}$ and $\ket{\uparrow\uparrow\!1}$ where we set $d\!\braket{\phi(t)|\uparrow\downarrow\!1}\!/dt=d\!\braket{\phi(t)|\uparrow\uparrow\!1}\!/dt=0$, we can compute $F_\pi$. Although by choosing $\Delta_B$ correctly, the unequal cavity coupling rates can be compensated, to minimize the gate time $T\propto (g_{\uparrow_A}g_{\downarrow_B})^{-1}$ it is optimal to have $g_{\uparrow_A}\simeq g_{\downarrow_B}=g$. In this case, we have
\begin{equation}
\begin{aligned}
    &F_\pi = \frac{1}{2}\left|\braket{e_2\uparrow 0|e^{-iT\hat{\mathcal{H}}_{\uparrow\uparrow}}|e_2\uparrow 0}-\braket{e_2\downarrow 0|e^{-iT\hat{\mathcal{H}}_{\uparrow\downarrow}}|e_2\downarrow 0}\right|\\
    &=\frac{1}{2}e^{-2\pi\Delta/C\kappa-\pi\kappa/2\Delta}\left|e^{i4\pi g^2/\Delta\delta_{eg}}\!+\!\cosh\!\left[\frac{\pi\kappa}{2\Delta}\right]e^{-i\pi\Delta_\epsilon\Delta/g^2}\right|.
\end{aligned}
\end{equation}
In the case where $\delta_{eg}\gg g^2/\Delta \gg \Delta_\epsilon$, this can be written as
\begin{equation}
\label{lukin-fidelity}
    F_\pi = e^{-2\pi\Delta/C\kappa-\pi\kappa/2\Delta}\cosh^2\!\!\left[\frac{\pi\kappa}{4\Delta}\right]\!+\!\mathcal{O}\left(\Delta_\epsilon^2,\delta_{eg}^{-2}\right).
\end{equation}
Then the total gate fidelity is given by $F_\text{gate} = (F_\pi + 1)/2$. This expression is maximized for the choice $2\Delta\simeq\kappa\sqrt{C}$ when $g< \kappa$ (see Figure 4 in the main manuscript). Then in the regime where $C\gg 1$, the maximum gate fidelity in the ideal regime can be expanded to the lowest order non-vanishing terms to acquire the result presented in equation (2) of the main text. 

If a photon is emitted during the gate and system A collapses to the ground state $\ket{\uparrow}_A$ prematurely, then the final pulse used to return system A coherently to the ground state will instead re-excite $\ket{\uparrow}_A$. Since system A is in an excited state after a failure, the fidelity $F_{\gamma\kappa}$ vanishes and so the non-Hermitian approximation is exact in this case.

\subsection{Raman virtual photon exchange}
The analysis of the Raman scheme follows similar to the simple virtual photon exchange scheme. However, with the addition of the Raman coupling there are two nested adiabatic processes occurring. To simplify the analysis, we will assume that any infidelity caused by pulses 1 and 2 used to shelve $\ket{\downarrow}_B$ are negligible compared to infidelity caused by the much slower Raman interaction.

For this scheme, we begin with two four-level systems coupled to a single cavity mode (see figure 5 of the main text). The general Hamiltonian that governs the evolution is $\hat{H} = \hat{H}_A + \hat{H}_B+\hat{H}_C + \hat{H}_I$. The quantum system $\hat{H}_k$ is given by
\begin{equation}
    \hat{H}_k = \omega_k\hat{\sigma}_{\uparrow_k}^\dagger\hat{\sigma}_{\uparrow_k}+\omega_{s_k}\hat{\sigma}_{s_k}^\dagger\hat{\sigma}_{s_k}-\omega_{g_k}\hat{\sigma}_{\uparrow\downarrow_k}^\dagger\hat{\sigma}_{\uparrow\downarrow_k},
\end{equation}
where $\omega_k$ is the frequency separation between the $\ket{\uparrow}_k$ and $\ket{e}_k$ states, $\omega_{s_k}$ is the separation of the shelving state $\ket{s}_k$ and $\ket{\uparrow}_k$, and $\omega_{g_k}$ is the separation of the $\ket{\uparrow}_k$ and $\ket{\downarrow}_k$ states. Also, $\hat{\sigma}_{\uparrow_k}\ket{e}_k=\ket{\uparrow}_k$, $\hat{\sigma}_{s_k}\ket{s}_k=\ket{\downarrow}_k$, and $\hat{\sigma}_{\uparrow\downarrow_k}\ket{\downarrow}_k=\ket{\uparrow}_k$ (see figure 5 of the main text). The cavity homogeneous evolution is $\hat{H}_C = \omega_C\hat{a}^\dagger\hat{a}$ for cavity frequency $\omega_C$, cavity photon anihilation (creation) operator $\hat{a}$ ($\hat{a}^\dagger$), and the interaction part is given by
\begin{equation}
    \hat{H}_I = \sum_{k\in{A,B}}g_k \hat{\sigma}_{\uparrow_k}^\dagger\hat{\sigma}_{\uparrow\downarrow_k}
    \hat{a} + \Omega_k \hat{\sigma}_{\uparrow_k}^\dagger e^{it\omega_{L_k}}+\text{h.c.},
\end{equation}
where $g_k$ is the cavity coupling rate of the $\ket{\uparrow}\rightarrow\ket{e}$ transition to the cavity mode and $\omega_{L_k}$ is the control laser frequency coupling $\ket{\downarrow}_k$ and $\ket{e}_k$ via the operator $\hat{\sigma}_{\uparrow_k}^\dagger\hat{\sigma}_{\uparrow\downarrow_k}\ket{\downarrow}_k = \hat{\sigma}_{\uparrow_k}^\dagger\ket{\uparrow}_k=\ket{e}_k$ with Rabi frequency $\Omega_k$. The dissipation is governed by the Lindblad master equation
\begin{equation}
    \dot{\rho} = -i[\hat{H},\hat{\rho}] +\kappa\mathcal{D}(\hat{a})\hat{\rho}+ \sum_{k}\gamma_k\mathcal{D}(\hat{\sigma}_{\uparrow_k})\hat{\rho}+\sum_{k}\gamma_{s_k}\mathcal{D}(\hat{\sigma}_{s_k})\hat{\rho},
\end{equation}
where $\mathcal{D}(\hat{A})\hat{\rho}=\hat{A}\hat{\rho}\hat{A}^\dagger -\{\hat{A}^\dagger\hat{A},\hat{\rho}\}/2$, $\kappa$ is the decay rate of the cavity photon, $\gamma_k$ is the decay rate of $\ket{e}_k$, and $\gamma_{s_k}$ as the decay rate of the shelving state $\ket{s}_k$. The corresponding effective non-Hermitian Hamiltonian is then
\begin{equation}
    \hat{\mathcal{H}}_\text{eff} = \hat{H} - \frac{i}{2}\left[\kappa\hat{a}^\dagger\hat{a}+ \gamma\sum_{k}\hat{\sigma}_{\uparrow_k}^\dagger \hat{\sigma}_{\uparrow_k}+\gamma_s\sum_k\hat{\sigma}_{s_k}^\dagger\hat{\sigma}_{s_k}\right]
\end{equation}
where we have assumed $\gamma_k=\gamma$ and $\gamma_{s_k}=\gamma_s$ for both quantum systems $A$ and $B$. Recall again that the effective Hamiltonian does not discriminate which ground state recycles the population. In effect, $\gamma_s$ represents the total decay rate out of $\ket{s}$ into any other state.

As with the virtual photon exchange, we can first break the system into four subsystems associated with the four basis states of the two-qubit space: $\{\ket{\uparrow\uparrow},\ket{\downarrow\uparrow},\ket{\uparrow\downarrow},\ket{\downarrow\downarrow}\}$. After shelving $\ket{\uparrow}_B$ to state $\ket{s}_B$, these states become $\{\ket{\uparrow\!s},\ket{\downarrow\!s},\ket{\uparrow\downarrow},\ket{\downarrow\downarrow}\}$. Then the driving fields $\Omega_A$ and $\Omega_B$ couple $\ket{\downarrow\uparrow}$ and $\ket{\uparrow\downarrow}$. Consequently, the fields also induce a phase on $\ket{\uparrow s}$ due to the AC Stark and cavity Lamb shifts. 
Since $\ket{s}_B$ is decoupled from the cavity and far-detuned from the driving fields, we only consider the dynamics in the subspaces affecting $\ket{\uparrow \downarrow}$ and $\ket{\uparrow \!s}$ dictated by $H_{\uparrow\downarrow}$ and $H_{\uparrow \uparrow}$. From the Hamiltonian, we can write $H_{\uparrow\downarrow}$ in the basis $\{\ket{\uparrow\downarrow\!0},\ket{e\!\downarrow \!0},\ket{\downarrow\downarrow\!1},\ket{\downarrow \!e\,0},\ket{\downarrow\uparrow \!0}\}$ as
\begin{equation}
\label{fengH1}
    H_{\uparrow \downarrow}= \begin{pmatrix}
    0 & \Omega_A & 0 & 0 & 0\\
    \Omega_A & \Delta_A & g_A & 0 & 0\\
    0 & g_A & -\delta_A & g_B & 0\\
    0 & 0 & g_B & \Delta_B + (\delta_B - \delta_A) & \Omega_B \\
    0 & 0 & 0 & \Omega_B & \delta_B - \delta_A
    \end{pmatrix},
\end{equation}
and $H_{\uparrow\uparrow}$ in the basis $\{\ket{\uparrow\!s\,0},\ket{e\,s\, 0},\ket{\downarrow\!s\,1}\}$ as
\begin{equation}
    H_{\uparrow\uparrow}=\begin{pmatrix}
    0 & \Omega_A & 0 \\
    \Omega_A & \Delta_A & g_A \\
    0 & g_A & -\delta_A
    \end{pmatrix},
\end{equation}
where $\Delta_k = \omega_k - \omega_{L_k}$ and $\delta_k = \omega_{L_k}+\omega_{g_k} - \omega_C$. The last index of each state indicates the photon number in the cavity mode. To obtain these time-independent subsystem Hamiltonians, we have moved into a rotating frame $\hat{H} \rightarrow e^{-it\hat{\mathcal{R}}}\hat{H}(t)e^{it\hat{\mathcal{R}}}-\hat{\mathcal{R}}$ defined by
\begin{equation}
\begin{aligned}
    \hat{\mathcal{R}}&=(\omega_C+\delta_B)\hat{a}^\dagger\hat{a} + \sum_k\omega_{L_k}\hat{\sigma}_{\uparrow_k}^\dagger\hat{\sigma}_{\uparrow_k}\\&+\sum_k\omega_{s_k}\hat{\sigma}_{s_k}^\dagger\hat{\sigma}_{s_k}+\omega_{g_B}\hat{\sigma}_{\uparrow\downarrow_B}^\dagger\hat{\sigma}_{\uparrow\downarrow_B}\\&+(\delta_A-\delta_B-\omega_{g_A})\hat{\sigma}_{\uparrow\downarrow_A}^\dagger\hat{\sigma}_{\uparrow\downarrow_A}.
\end{aligned}
\end{equation}
This rotating frame preserves the desired relative phase evolution between $\ket{\uparrow\downarrow}$ and $\ket{\uparrow\uparrow}$ because it is defined using local operators only.

The evolution in remaining subsystems can be neglected: $H_{\downarrow\uparrow}=0$ and $H_{\downarrow\downarrow}=0$ for the states that are not coupled to the driving fields. Similar to the prevous scheme, the total gate fidelity can then be simplified to $F_\text{gate} = (F_\pi + 1)/2$ where $F_\pi=|\bra{\phi(T)}(\ket{\uparrow\uparrow}-\ket{\uparrow\downarrow})|/\sqrt{2}$ for initial state $\ket{\psi(0)}=(\ket{\uparrow\uparrow}+\ket{\uparrow\downarrow})/\sqrt{2}$.

At the two-photon resonance ($\delta_A=\delta_B=\delta$), $H_{\uparrow\downarrow}$ will perform a $\pi$-phase rotation on $\ket{\uparrow\downarrow}$. However, unlike the the simple exchange scheme, the Raman exchange scheme can operate when $\Delta_A$ and $\Delta_B$ are not restricted to a fixed relation, allowing for a gate to be performed between two quantum systems that have unequal optical transitions. However, a $\pi$ phase can only be achieved when $\Omega_B$ is selected to compensate for $g_A\neq g_B$ and $\Delta_A\neq\Delta_B$. By adiabatically eliminating the excited state and cavity amplitudes where we set $d\!\braket{\phi(t)|e\!\downarrow\!0}\!/dt=d\!\braket{\phi(t)|\!\downarrow\downarrow\!1}\!/dt=d\!\braket{\phi(t)|\!\downarrow\!e~0}\!/dt=0$, the optimal Rabi frequency relation is found to be
\begin{equation}
    \Omega_B = \Omega_A\sqrt{\frac{g_A^2+\delta\Delta_A}{g_B^2+\delta\Delta_B}}\simeq\Omega_A\sqrt{\frac{\Delta_A}{\Delta_B}},
\end{equation}
corresponding to the time required to achieve a $\pi$ phase of
\begin{equation}
    T = \pi\frac{g_A^2\Delta_A+g_B^2\Delta_B+\delta\Delta_A\Delta_B}{g_A g_B\Omega_A\Omega_B}\simeq\pi\frac{\delta\Delta_A\Delta_B}{g_A g_B\Omega_A\Omega_B},
\end{equation}
assuming $g_k^2\ll\delta\Delta_k$.

The non-Hermitian parts are given by corresponding decay rates of each state amplitude
\begin{equation}
\begin{aligned}
    \hat{\mathcal{H}}_{\uparrow\downarrow}&=\hat{H}_{\uparrow\downarrow}\\&-\frac{i}{2}\left(\gamma\ket{e\!\downarrow\!0}\bra{e\!\downarrow\!0} + \kappa\ket{\downarrow\downarrow\!1}\bra{\downarrow\downarrow\!1} + \gamma\ket{\downarrow\!e \,0}\bra{\downarrow\!e\,0}\right)\\
    \hat{\mathcal{H}}_{\uparrow\uparrow}&=\hat{H}_{\uparrow\uparrow}-\frac{i}{2}\left(\gamma\ket{e\,s\,0}\bra{e\,s\, 0}+\kappa\ket{\uparrow\!s\, 1}\bra{\uparrow\!s\,1}\right)
\end{aligned}
\end{equation}
where we have assumed that $\gamma_s\ll \gamma,\kappa$.
We analyze the fidelity in the case where $g_A=g_B=g$, $\delta_A\simeq\delta_B\simeq\delta$ but $\delta_\epsilon=|\delta_A-\delta_B|\ll \delta$ is nonzero, $\Delta_A\simeq\Delta_B\simeq\Delta$ but $\Delta_\epsilon=|\Delta_A\!-\!\Delta_B|\ll \Delta$ is nonzero, and also $\Omega_A\simeq\Omega_B\sqrt{\Delta_B/\Delta_A}\simeq\Omega$. Under these conditions, and after adiabatically eliminating the state amplitudes that are only virtually populated, the fidelity overlap is found to be
\begin{equation}
\label{pifid}
\begin{aligned}
    F_\pi&=|\bra{\uparrow\!s\, 0}e^{-i\mathcal{H}_{\uparrow\uparrow}T}\ket{\uparrow\!s\,0}-\bra{\uparrow\downarrow\!0}e^{-i\mathcal{H}_{\uparrow\downarrow}T}\ket{\uparrow\downarrow\!0}|\\
    &\simeq  e^{-2\pi\delta/C\kappa-\pi\kappa/2\delta}\cosh^2\left[\frac{\pi\kappa}{4\delta}\right]
    +\mathcal{O}\left(\Delta_\epsilon^2,\delta_\epsilon^2,\gamma_s\right),
\end{aligned}
\end{equation}
where we assume that $C\gg 1$. Notice that this solution mirrors that of the previous scheme but now the two-photon detuning $\delta$ plays the same role that the cavity detuning $\Delta$ did prior. 
The maximum cooperativity-limited fidelity is then given when $2\delta=\kappa\sqrt{C}$. In the regime where $C\gg 1$ and $\gamma_s\ll \gamma$, the maximum fidelity can be expressed as
\begin{equation}
\label{fidelity-feng}
    \!F_\text{max}\!= 1 - \frac{\pi}{\sqrt{C}}-\frac{\pi^2}{16}\left[\frac{T_o^2\delta_\epsilon^2}{4\pi^2}+\frac{\Delta_\epsilon^2}{\Delta^2}-\frac{18}{C}\right]
\end{equation}
at the optimal gate time of $T_o = (\Delta/\Omega)^2(2\pi/\gamma\sqrt{C})$.

The maximum fidelity given by equation (\ref{fidelity-feng}) relies on the satisfaction of adiabatic criteria. Unlike the previous scheme where $C\gg 1$ and $g\leq \kappa$ was enough to ensure adiabatic evolution in the ideal detuning regime, the Raman process places additional constraints on the driving field parameters to achieve adiabatic evolution. Primarily, it is necessary for $\Omega\ll \Delta$. However, the magnitudes of $\Delta$ and $\Omega$ are also limited by other system parameters. The lower bound on $\Delta$ for a given $\Omega/\Delta$ ratio can be determined by considering the regime $\delta\Delta< g^2$ where cavity Rabi oscillations cause infidelity. This limit can be solved by adiabatically eliminating both the excited state amplitudes and the cavity mode amplitude. Likewise, the upper bound on $\Delta$ for a given $\Omega/\Delta$ ratio can be determined by considering the regime $\delta\Delta< \Omega^2$ where Rabi oscillations induced by the driving field cause infidelity. These limits can be analytically solved by adiabatically eliminating only the excited state amplitudes.

In the ideal regime where $\gamma_s/\gamma\ll 2(\Omega/\Delta)^2$, $\Delta_A\simeq \Delta_B$, and $\delta_\epsilon\ll 2\pi T_o^{-1}$, the total gate fidelity is well-approximated by
\begin{equation}
\label{bounds}
\begin{aligned}
    F_\text{gate}\simeq & \frac{1}{2}\!\left(\cos^2\!\left[\frac{\pi\Omega}{4\Delta}\right]\cos^2\!\left[\frac{\pi\Omega^2}{2\delta\Delta}\right]\sin\!\left[\frac{\pi/2}{1\!+\!g^2/\delta\Delta}\right]\!F_\pi\!+1\!\right)\\
    &\simeq F_\text{max}- \frac{\pi^2}{16}\left(\frac{\Omega^2}{2\Delta^2}+\frac{2\Omega^4}{\delta^2\Delta^2}+\frac{g^4}{\delta^2\Delta^2}\right)
\end{aligned}
\end{equation}
where the additional $-\Gamma T_o$ scaling can be added to account for decoherence infidelity. These extra constraints on the adiabatic evolution are independent of the cavity cooperativity but they do place bounds on the regime where the fidelity is only limited by the cavity cooperativity. In addition, they will place bounds on the gate time. Combining the results from equations (\ref{pifid}) and (\ref{bounds}) provides a very accurate estimate of the fidelity given by the non-Hermitian Hamiltonian, as shown by the black dashed line in figure \ref{fig:ramanscheme} of the main text.

The upper bound on $\Delta$ will dictate the maximum possible spectral separation of the optical transitions. In turn, for a fixed $\Omega/\Delta$ ratio, $\Delta$ is limited by the condition $\pi\Delta < \delta(\Delta/\Omega)^2$ needed to maintain adiabatic evolution and the ratio $\Omega/\Delta$ itself is limited from below by decoherence due to the gate time scaling of $T\propto(\Delta/\Omega)^2$. To solve for the minimum $\Delta/\Omega$ and maximum $\Delta$ corresponding to the maximum $\Delta_\epsilon$, we optimize the expression
\begin{equation}
    \frac{2\pi\Gamma\Delta^2}{\gamma\Omega^2\sqrt{C}} + \frac{\pi^2\Omega^4}{8\delta^2\Delta^2}+\frac{\pi^2\Delta_\epsilon^2}{16\Delta^2}= \frac{\pi}{\sqrt{C}}
\end{equation}
when $2\delta=\kappa\sqrt{C}$, where the first term is the infidelity $\Gamma T_o$ due to the effective decoherence $\Gamma$, the second term captures the condition $\Delta> \Omega$ from equation (\ref{bounds}) to maintain adiabatic evolution, and the third term captures the infidelity from equation (\ref{fidelity-feng}) due to spectral separation of the optical transitions. From this expression, we find that the maximum spectral separation that will result in an infidelity less than the cooperativity-limited value is
\begin{equation}
    \Delta_\epsilon= \frac{\kappa\gamma}{\pi\Gamma\sqrt{8}},
\end{equation}
which corresponds to $\Omega/\Delta=2\sqrt{\Gamma/\gamma}$ and $2\Delta=\Delta_\epsilon\sqrt{\pi\sqrt{C}}$. 
This implies that $\Delta_A/\Delta_B\simeq 1 \pm \sqrt{2/\pi\sqrt{C}}$, which validates the initial assumption that $\Delta_\epsilon\ll\Delta$. The corresponding gate time for these conditions is $T=\pi/(2\Gamma\sqrt{C})$.

In the adiabatic regime, the emission of a photon will collapse the system into the mixed state $\hat{\rho}_{\gamma\kappa}\simeq(1/2)\ket{\downarrow}\!\bra{\downarrow}_A\otimes(\ket{\downarrow}\!\bra{\downarrow}+\ket{\text{s}}\!\bra{\text{s}})_B$. Hence the fidelity for a failure is $F_{\gamma\kappa}= 1/2$. This implies that the maximum error in the non-Hermitian solution $F_\text{gate}$ given above is \mbox{$\sqrt{F_\text{gate}^2+(1-F_\text{gate}^2)/4}-F_\text{gate}\sim \pi/4\sqrt{C}$}. For example, where the non-Hermitian approximation gives 0.9 (0.99), the true fidelity from the full master equation is not more than 0.93 (0.993).

\bibliography{ref}

\end{document}